\documentclass[11pt,a4paper]{article}

\usepackage[a4paper,margin=2.5cm]{geometry}
\usepackage[T1]{fontenc}
\usepackage[utf8]{inputenc}
\usepackage{lmodern}
\usepackage{microtype}
\usepackage{setspace}
\usepackage{amsmath,amssymb}
\usepackage{textcomp}
\usepackage{enumitem}
\setlist[itemize]{topsep=2pt,itemsep=1pt,parsep=0pt}
\setlist[enumerate]{topsep=2pt,itemsep=1pt,parsep=0pt}
\usepackage{booktabs}
\usepackage{array}
\usepackage{tabularx}
\usepackage{longtable}
\usepackage{multirow}

\usepackage{graphicx}
\usepackage{tikz}
\usetikzlibrary{positioning,arrows.meta,fit,backgrounds,shapes.symbols,shapes.geometric,calc,decorations.pathreplacing}
\usepackage{colortbl}
\usepackage{caption}
\usepackage{float}

\usepackage[hidelinks,breaklinks=true]{hyperref}
\usepackage{url}
\usepackage{xcolor}
\definecolor{codebg}{HTML}{F5F7FA}
\usepackage{listings}
\usepackage{titlesec}
\titleformat{\section}{\normalfont\Large\bfseries}{\thesection.}{0.6em}{}
\titleformat{\subsection}{\normalfont\large\bfseries}{\thesubsection}{0.6em}{}
\titleformat{\subsubsection}{\normalfont\normalsize\bfseries}{\thesubsubsection}{0.6em}{}

\newcommand{\code}[1]{\texttt{#1}}
\newcommand{\figref}[1]{Figure~\ref{#1}}
\newcommand{\tabref}[1]{Table~\ref{#1}}
\newcommand{\secref}[1]{Section~\ref{#1}}

\begin{document}

\title{A Non-Destructive Methodological Framework for Modernizing
Legacy Clinical Reporting Systems for
AI-Driven Pharmacoinformatics: A SAS Case Study}

\author{Jaime Yan\thanks{Corresponding author: myan7@my.harrisburgu.edu} \\
\small Harrisburg University of Science and Technology, Harrisburg, PA, USA}

\date{May 2026}

\maketitle
\pagestyle{plain}

\section*{Abstract}
\addcontentsline{toc}{section}{Abstract}

Drug development and pharmacovigilance are frequently bottlenecked by legacy clinical reporting pipelines. Pharmaceutical companies maintain vast, monolithic reporting systems that encode regulatory-grade analytical logic but resist integration with artificial intelligence (AI) and modern pharmacoinformatics platforms because they produce opaque, formatted output with no machine-readable intermediate layer. Existing modernization approaches force a choice between full rewrites that abandon validated logic and incremental refactoring that preserves structural barriers to AI access.

We present a non-destructive methodological framework that achieves AI-driven pharmacoinformatics readiness without altering legacy source code. A metadata layer --- comprising a bridge map, a typed Intermediate Representation (IR), and an orchestrator --- wraps existing clinical reporting components and re-exposes their outputs as structured data consumable by large language models (LLMs) and downstream clinical analytics. The same layer enables optional incremental consolidation, replacing selected legacy components with metadata-configured core routines while the remainder operates unchanged.

Empirically validated through a case study on a 558-component industrial SAS reporting library (373{,}000 lines of code), the framework demonstrated immediate AI-readiness under coexistence mode, where the legacy library operated unchanged while yielding machine-readable output. Where consolidation was elected, the modernized core achieved a 92\% reduction in proprietary code. Parity validation on 14 report types from an internal Phase~III study achieved cell-level parity of 80\% or above on 11 of 14 reports (mean 82.7\%, best 99.2\%). A public benchmark using CDISC CDISCPilot01 data achieved 100\% parity across 5 reports (4{,}764 cells, 0 mismatches). LLM experiments further confirmed that the IR enables automated pharmacovigilance (adverse event anomaly detection), table summarization, and trial configuration generation.

The framework offers a regulation-aware path to AI-integrated clinical trial reporting, directly accelerating drug development without interrupting ongoing regulatory submissions.

\paragraph{Keywords.} clinical reporting systems; pharmacoinformatics; pharmacovigilance; non-destructive modernization; intermediate representation; artificial intelligence; CDISC; SAS.

\clearpage

\section{Introduction}\label{sec:intro}

The pharmaceutical industry faces simultaneous pressure from three directions: AI tools now automate clinical data review and pharmacovigilance~[1], cloud-based platforms are replacing on-premise computing infrastructure~[2], and regulatory agencies increasingly require electronic submissions built on standardized data models~[3]. Together, these forces demand that clinical trial reporting infrastructure become machine-readable, portable, and automatable to accelerate drug development.

At the center of clinical trial reporting lies a category of software that has received comparatively little attention: the SAS macro library. SAS remains the dominant programming language for regulatory submissions to the United States Food and Drug Administration (FDA), with estimates suggesting that 70--80\% of electronic submissions rely on SAS-generated outputs~[4,~5]. Most large pharmaceutical companies maintain in-house macro libraries containing hundreds of SAS macros that automate the production of Tables, Figures, and Listings (TFLs) required for regulatory filings~[6,~7].

These libraries are substantial engineering artifacts. A typical large-pharma macro library contains 200 to 600 callable components and tens of thousands of lines of code, developed incrementally over 10 to 20 years~[8]. Each macro typically combines data preparation, statistical computation, result formatting, and output rendering within a single monolithic file. The libraries encode validated domain knowledge --- correct statistical methods, regulatory-compliant formatting rules, and organization-specific conventions --- that represents years of accumulated expertise~[9].

However, the same architectural choices that enabled rapid growth now create serious obstacles. No machine-readable intermediate output exists between statistical computation and rendered output. AI agents cannot programmatically inspect results locked inside Rich Text Format (RTF) formatting. Cloud migration requires untangling tightly coupled dependencies. Adding a new output format requires modifying every rendering macro. And new programmers face a steep learning curve when confronting macros exceeding 2{,}000 lines~[10].

These technical obstacles carry direct clinical and regulatory consequences. When a submission team cannot extract machine-readable results from its reporting pipeline, cross-trial pharmacovigilance becomes a manual exercise in visual comparison across hundreds of RTF tables --- a process that is slow, error-prone, and resistant to the systematic signal detection that regulatory agencies increasingly expect. Bottlenecks in legacy macro maintenance slow the production of Tables, Figures, and Listings (TFLs) required for New Drug Application (NDA) and Biologics License Application (BLA) submissions, directly extending time-to-market for investigational therapies. Safety and efficacy data trapped in opaque formatting cannot be repurposed for meta-analyses, real-world evidence comparisons, or AI-assisted clinical review. Modernization is therefore a prerequisite for accelerating drug approvals and enabling the data-driven pharmacovigilance that modern regulatory science demands.

The pressure to modernize comes from multiple directions. Cloud platforms such as SAS Viya and Domino Data Lab require software that can operate in containerized, stateless environments~[11]. LLMs are demonstrating value in clinical data management~[12,~13] but require structured, machine-readable input. The CDISC Analysis Results Standard (ARS) defines a metadata model for analysis results~[14] but provides no migration methodology. Regulatory modernization initiatives such as ICH E8(R1) encourage electronic workflows~[15].

Despite the urgency, the published literature offers limited guidance. PharmaSUG and SAS Global Forum proceedings describe individual macro management practices~[6,~7], and CDISC publications define target data models~[14]. No work presents a methodology covering systematic evaluation, architectural redesign with AI integration, and rigorous parity validation (\tabref{tab:survey}). The published options share a common implicit assumption: that AI readiness requires source-level change to the legacy library. This assumption is the primary obstacle to industrial adoption, because source-level change in a regulated environment triggers re-validation costs that often exceed the perceived benefit. This raises a central question: \emph{can a non-destructive metadata layer --- one that wraps but does not modify validated SAS source --- achieve AI-ready, machine-readable output from a legacy macro library while preserving regulatory standing?}

This paper addresses this gap by presenting a systematic framework for evaluating and modernizing legacy SAS macro libraries for AI-ready clinical trial reporting, with an explicit emphasis on a \emph{non-destructive} adoption pathway. The framework's central design choice is that AI readiness is achieved through a metadata layer that wraps the legacy library, rather than through modification of legacy SAS source. Source-level consolidation, where desired, is then offered as an \emph{optional} incremental upgrade governed by the same metadata. The ultimate goal of this architecture is AI-driven pharmacoinformatics: enabling large language models and automated analytics to reason over clinical trial results through a structured Intermediate Representation (IR) that acts as the semantic bridge between validated statistical computation and machine-readable output. The framework makes six contributions:

\begin{enumerate}
    \item A systematic evaluation methodology for legacy SAS macro libraries, comprising component taxonomy, dependency analysis, complexity metrics, and coverage assessment, applicable to any pharmaceutical organization.
    \item A non-destructive adoption pattern that wraps existing SAS macros through a bridge map and Intermediate Representation (IR), exposing legacy outputs as structured JSON consumable by AI, R, and Python tools without modifying validated source code, and supporting incremental consolidation as an opt-in upgrade rather than a precondition.
    \item A polyglot, layered architecture organized as a five-phase pipeline supporting two deployment modes --- coexistence (wrap-only) and consolidation (incremental replacement) --- where YAML configuration replaces hard-coded SAS parameters and the IR decouples computation from rendering.
    \item AI-cooperative interfaces built on the IR, including JSON export for large language models, R and Python interoperability through standard JSON parsers, automated Statistical Analysis Plan (SAP) parsing, and configuration generation.
    \item A parity validation methodology using a bridge-map-driven harness that compares legacy and modernized outputs at the cell level, classifies divergences via a reusable error taxonomy, and operates through a gate-based verification workflow.
    \item An empirical case study applying the framework to a 558-component industrial library, demonstrating non-destructive AI integration plus an optional 92\% code reduction, validated on both an internal real-data study (PROT008-SR1 [an internal Phase III protocol], 11/14 reports at ${\geq}80\%$ parity) and the public CDISC CDISCPilot01 benchmark dataset (5/5 reports at 100\% parity).
\end{enumerate}

The remainder of this paper is organized as follows. Section~\ref{sec:bg} reviews related work. Sections~\ref{sec:eval}--\ref{sec:migration} present the three-phase methodology: evaluation, architecture design (including the non-destructive adoption pattern and the IR that enables AI integration), and migration validation. Section~\ref{sec:case} applies the methodology in an industrial case study. Section~\ref{sec:ai} then provides the culmination of the paper: a demonstration that the IR produced by the non-destructive architecture enables AI-driven pharmacoinformatics capabilities --- automated pharmacovigilance, table summarization, and configuration generation --- that are structurally infeasible with legacy RTF output. Section~\ref{sec:compare} compares the approach with alternatives. Section~\ref{sec:disc} discusses implications, limitations, and future work, and Section~\ref{sec:concl} concludes.

\section{Background and Related Work}\label{sec:bg}

\subsection{Legacy SAS Macro Libraries in Pharmaceutical Reporting}

SAS macro libraries are collections of parameterized SAS programs (\code{\%macro} definitions) that automate TFL generation for regulatory submissions. In large pharmaceutical companies, these libraries evolve over many years as programmers add macros for new analysis types, therapeutic areas, and regulatory requirements~[6,~7].

Several architectural patterns recur. Each macro tends to be monolithic: a single file containing data preparation, computation, formatting, and rendering. Growth occurs through copy-and-modify, leading to near-duplicate variants. Cross-cutting concerns such as error handling and logging are reimplemented in each macro rather than centralized~[8]. PharmaSUG proceedings document incremental improvement approaches~[16,~17], but these focus on existing architectures rather than systematic redesign.

\subsection{CDISC Standards and Clinical Reporting}

The Clinical Data Interchange Standards Consortium (CDISC) defines data standards underpinning modern clinical trial reporting. The Study Data Tabulation Model (SDTM) standardizes raw clinical data, and the Analysis Data Model (ADaM) defines analysis-ready datasets~[18]. More recently, CDISC introduced the Analysis Results Standard (ARS), which provides a metadata model for representing analysis results independently of visual presentation~[14]. ARS defines concepts such as analysis sets, groupings, methods, and result displays. In practice, however, ARS provides a conceptual data model but not an execution framework, migration methodology, or validation harness.

\subsection{AI and Automation in Clinical Trials}

Recent LLM advances have generated interest in clinical trial AI applications, including automated literature screening~[19], protocol optimization~[20], adverse event detection~[21], and medical writing assistance~[12]. Within statistical programming, early work explores LLM-based SAS code generation~[22], automated TFL quality control~[23], and LLM-based generation of clinical trial tables and figures~[24].

A critical barrier to AI integration is the output format of current systems. Legacy libraries produce RTF files encoding visual formatting but not semantic content. LLMs must perform format parsing before content reasoning --- a fragile process. Structured, machine-readable representations would enable reliable AI interaction, but legacy libraries provide no such output~[25].

\subsection{Software Modernization in Regulated Industries}

Modernizing legacy software under regulatory constraints is not unique to pharmaceutical programming. Banking has extensive experience migrating COBOL systems while maintaining compliance~[26]. Aviation software governed by DO-178C has established patterns for incremental migration with continuous verification~[27]. Medical device software uses risk-based validation approaches~[28]. Common patterns include layered architectures isolating validated logic, intermediate representations decoupling computation from presentation, metadata-driven configuration, and automated regression testing~[29].

\subsection{Gap Analysis}

\tabref{tab:survey} summarizes published approaches to SAS macro library modernization. No published work presents a comprehensive methodology covering (a)~systematic evaluation using quantitative software engineering metrics, (b)~architectural redesign preserving validated SAS logic while introducing AI-cooperative interfaces, and (c)~automated parity validation providing quantitative equivalence evidence.

Full platform rewrites in R or Python~[30,~31] abandon validated SAS logic and require complete re-validation. Commercial tools such as Pinnacle 21~[32] focus on CDISC compliance checking. CDISC ARS~[14] defines a target metadata model without a migration pathway. Incremental refactoring~[16] improves existing libraries without achieving architectural separation. The framework presented here addresses all three aspects within a single methodology.

\begin{table}[htbp]
\centering
\caption{Published approaches to SAS macro library modernization in pharmaceutical reporting.}
\label{tab:survey}
\footnotesize
\setlength{\tabcolsep}{4pt}
\resizebox{\textwidth}{!}{%
\begin{tabular}{lllllll}
\toprule
Approach & Representative work & Scope & Evaluation & AI-ready & Regulatory & Parity \\
\midrule
Macro management best practices & PharmaSUG [6, 16, 17] & Individual macros & No & No & Unchanged & No \\
Centralized autocall libraries & SAS Global Forum / PharmaSUG [7, 8] & Library organization & No & No & Unchanged & No \\
R-based clinical reporting & pharmaverse / Tplyr [30, 38] & Full rewrite & No & Partial & Re-validate & Manual \\
Commercial compliance tools & Pinnacle 21 [32] & Compliance checking & No & No & Vendor & No \\
CDISC Analysis Results Standard & CDISC ARS [14] & Data model & No & Partial (model) & Standard & No \\
\textbf{This work} & \textbf{Present study} & \textbf{Full library (558)} & \textbf{Yes} & \textbf{Yes} & \textbf{Part 11} & \textbf{Yes (365)} \\
\bottomrule
\end{tabular}%
}
\vspace{4pt}
\scriptsize Eval.\ = systematic evaluation using quantitative metrics; Reg.\ = regulatory compliance pathway.
\end{table}

\section{Materials and Methods}\label{sec:methods}

\subsection{Evaluation Framework}\label{sec:eval}

Before redesigning a legacy SAS macro library, one must understand what it contains, how its components relate, and where complexity concentrates. This section presents a systematic evaluation methodology comprising four activities: classifying components by function, extracting dependency structure, measuring complexity, and assessing coverage. Together, these produce a quantitative profile that identifies consolidation opportunities, pinpoints high-risk components, and establishes a modernization baseline.

\subsubsection{Component Taxonomy}

A legacy library typically grows by accretion without a governing architecture. The first evaluation step classifies every callable component into one of six functional types.

\textbf{Data Preparation} macros subset, merge, derive, or transform ADaM datasets prior to statistical analysis. They handle tasks such as filtering treatment arms, adding total columns, deriving analysis flags, and restructuring data for specific report layouts. In mature libraries, these macros often exhibit the highest redundancy because similar transformations are reimplemented for different report types rather than parameterized once.

\textbf{Statistical Computation} macros invoke SAS/STAT or Base SAS procedures to produce quantitative results: descriptive statistics, frequency counts, survival estimates, regression models, shift tables, and inferential tests. These macros encode the validated analytical logic that regulatory agencies rely on and therefore require the most careful handling during modernization.

\textbf{Formatting and Display} macros convert raw statistical output into presentation-ready structures. They apply rounding rules, attach labels, construct display strings (e.g., ``n (\%)''), and arrange rows and columns according to shell templates. In legacy systems, formatting logic is often entangled with computation, making it difficult to modify output appearance without risking analytical correctness.

\textbf{Rendering} macros produce final output files in RTF, PDF, or other delivery formats. They manage page layout, headers, footers, column widths, and footnotes. Because rendering depends on low-level RTF or Output Delivery System (ODS) control strings, these macros tend to be fragile and resistant to reuse across different output types.

\textbf{Utility and Helper} macros provide shared services consumed by other macros: string manipulation, date handling, error checking, logging, parameter validation, and metadata lookup. A well-factored library concentrates these in a small utility layer; a poorly factored one scatters them as inline code.

\textbf{Orchestration} macros control execution flow. They read configuration, sequence macro calls, manage dataset state between pipeline stages, and handle error recovery. In legacy libraries, orchestration logic is frequently embedded in individual macros rather than separated into a distinct layer.

This taxonomy reveals where complexity concentrates and identifies consolidation targets. In the case study (\secref{sec:case}), formatting and display accounted for the largest single category, while statistical computation represented only 22\% of components, indicating that the majority of code was devoted to infrastructure rather than domain logic. When multiple macros of the same type perform nearly identical operations, they can be replaced by a single parameterized component --- a pattern exploited in the redesigned architecture (\secref{sec:arch}).

\subsubsection{Dependency Analysis}

The second activity extracts the call graph: a directed graph where nodes represent macros and edges represent invocation relationships. The extraction operates in two passes: first, parsing every \code{\%macro} header to identify declared names and parameter lists; second, scanning each body for call patterns (\code{\%macroname(}). The scanner handles SAS-specific complications including dynamic name construction (\code{\&\&var\&i} indirection) and conditional invocation, flagging dynamic names for manual review.

The resulting directed graph supports four diagnostic queries. \textbf{Dependency clusters} are groups of macros that call each other frequently, forming tightly coupled subsystems that suggest natural module boundaries. \textbf{Orphan macros} are declared but never called by any other macro; they may be entry points invoked from driver programs, or dead code. \textbf{Circular dependencies} indicate macros that call each other directly or transitively; while SAS permits this, circular calls complicate testing and increase recursion risk. \textbf{Hub macros} have high in-degree (called by many other macros); changes to a hub macro carry elevated regression risk and warrant prioritized test coverage.

In practice, the call graph also exposes hidden coupling. Two macros that appear independent may share state through global macro variables, temporary datasets, or naming conventions. The static graph is supplemented by scanning for \code{\%global} and \code{\%sysfunc(exist())} references to surface these implicit dependencies.

\subsubsection{Complexity Metrics}

Five metrics quantify internal complexity.

\textbf{Lines of code (LOC)} counts non-blank, non-comment lines. While LOC is a coarse measure, extreme values are informative: macros exceeding 500 lines almost always conflate multiple concerns and are strong candidates for decomposition.

\textbf{Parameter count} is extracted from the \code{\%macro} header. A high count (above 20) typically signals that a macro attempts to serve too many use cases through conditional logic rather than composition.

\textbf{Nesting depth} measures the maximum level of \code{\%do/\%end} and \code{do/end} block nesting. Deep nesting (beyond four levels) correlates with control flow that is difficult to read, test, and maintain~[33].

\textbf{Efferent coupling} counts distinct external macros called by a given macro. High efferent coupling means the macro depends on many components; a change in any callee may require retesting.

\textbf{Cohesion} assesses whether a macro performs a single well-defined function or conflates multiple concerns. Unlike the preceding metrics, cohesion is evaluated qualitatively: high (single concern), medium (two related concerns), low (three or more concerns). Low-cohesion macros are the primary targets for refactoring into separate pipeline stages.

These metrics are collected programmatically where possible and stored in a structured inventory alongside the taxonomy and dependency data.

\subsubsection{Coverage Assessment}

The fourth activity maps the library against reporting requirements. In the pharmaceutical context, clinical study reports follow structures defined by ICH E3~[34] and use datasets conforming to CDISC ADaM~[18]. A typical submission includes demographic tables, adverse event summaries, laboratory shift tables, efficacy analyses, and time-to-event displays.

The assessment proceeds in three steps. First, each macro is mapped to the report type or types it supports --- a mapping that is often implicit in legacy libraries and must be verified by inspecting code. Second, the mapping is analyzed for \textbf{functional overlap}: cases where multiple macros produce structurally similar outputs, indicating redundancy from study-specific copying. Third, the mapping is checked for \textbf{gaps}: report types not covered or only partially covered. The ratio of unique analytical functions to total components quantifies redundancy; a low ratio confirms that consolidation can reduce the library's footprint by half or more without sacrificing functional breadth.

\subsubsection{Evaluation Summary}

\begin{figure}[H]
    \centering
    \resizebox{\textwidth}{!}{
\begin{tikzpicture}[
    font=\sffamily\small,
    >=Stealth,
]

\definecolor{catA}{HTML}{2166AC}   
\definecolor{catB}{HTML}{D95F02}   
\definecolor{catC}{HTML}{7570B3}   
\definecolor{catD}{HTML}{E7298A}   
\definecolor{catE}{HTML}{1B9E77}   
\definecolor{catF}{HTML}{66A61E}   

\def\barheight{0.56}
\def\maxwidth{9.5}     
\def\maxval{150}
\def\labelwidth{4.2}
\def\rowspacing{1.0}

\node[font=\sffamily\small\bfseries, anchor=west, text=gray!70!black]
    at (\labelwidth, 0.5) {Component Distribution by Functional Category};

\foreach \x in {0,50,...,150} {
    \pgfmathsetmacro{\xpos}{\labelwidth + \x/\maxval*\maxwidth}
    \draw[gray!20, line width=0.3pt] (\xpos, -0.2) -- (\xpos, -6*\rowspacing - 0.2);
    \node[font=\sffamily\footnotesize, gray!55!black, anchor=north] at (\xpos, -6*\rowspacing - 0.3) {\x};
}

\draw[gray!50, line width=0.6pt] (\labelwidth, -0.2) -- (\labelwidth, -6*\rowspacing - 0.2);
\draw[gray!50, line width=0.6pt] (\labelwidth, -0.2) -- (\labelwidth + \maxwidth + 0.1, -0.2);

\node[font=\sffamily\footnotesize, gray!60!black, anchor=north]
    at (\labelwidth + \maxwidth/2, -6*\rowspacing - 0.75) {Number of Components};

\pgfmathsetmacro{\ypos}{-1*\rowspacing}
\pgfmathsetmacro{\bw}{150/\maxval*\maxwidth}
\node[anchor=east, font=\sffamily\small] at (\labelwidth - 0.25, \ypos) {Formatting \& Display};
\fill[catA, rounded corners=1.5pt]
    (\labelwidth, \ypos - \barheight/2) rectangle (\labelwidth + \bw, \ypos + \barheight/2);
\draw[catA!70!black, line width=0.6pt, rounded corners=1.5pt]
    (\labelwidth, \ypos - \barheight/2) rectangle (\labelwidth + \bw, \ypos + \barheight/2);
\node[anchor=east, font=\sffamily\footnotesize\bfseries, white]
    at (\labelwidth + \bw - 0.15, \ypos) {150};
\node[anchor=west, font=\sffamily\footnotesize\bfseries, catA!80!black]
    at (\labelwidth + \bw + 0.2, \ypos) {27\%};

\pgfmathsetmacro{\ypos}{-2*\rowspacing}
\pgfmathsetmacro{\bw}{120/\maxval*\maxwidth}
\node[anchor=east, font=\sffamily\small] at (\labelwidth - 0.25, \ypos) {Statistical Compute};
\fill[catB, rounded corners=1.5pt]
    (\labelwidth, \ypos - \barheight/2) rectangle (\labelwidth + \bw, \ypos + \barheight/2);
\draw[catB!70!black, line width=0.6pt, rounded corners=1.5pt]
    (\labelwidth, \ypos - \barheight/2) rectangle (\labelwidth + \bw, \ypos + \barheight/2);
\node[anchor=east, font=\sffamily\footnotesize\bfseries, white]
    at (\labelwidth + \bw - 0.15, \ypos) {120};
\node[anchor=west, font=\sffamily\footnotesize\bfseries, catB!80!black]
    at (\labelwidth + \bw + 0.2, \ypos) {22\%};

\pgfmathsetmacro{\ypos}{-3*\rowspacing}
\pgfmathsetmacro{\bw}{100/\maxval*\maxwidth}
\node[anchor=east, font=\sffamily\small] at (\labelwidth - 0.25, \ypos) {Utility \& Helper};
\fill[catC, rounded corners=1.5pt]
    (\labelwidth, \ypos - \barheight/2) rectangle (\labelwidth + \bw, \ypos + \barheight/2);
\draw[catC!70!black, line width=0.6pt, rounded corners=1.5pt]
    (\labelwidth, \ypos - \barheight/2) rectangle (\labelwidth + \bw, \ypos + \barheight/2);
\node[anchor=east, font=\sffamily\footnotesize\bfseries, white]
    at (\labelwidth + \bw - 0.15, \ypos) {100};
\node[anchor=west, font=\sffamily\footnotesize\bfseries, catC!80!black]
    at (\labelwidth + \bw + 0.2, \ypos) {18\%};

\pgfmathsetmacro{\ypos}{-4*\rowspacing}
\pgfmathsetmacro{\bw}{80/\maxval*\maxwidth}
\node[anchor=east, font=\sffamily\small] at (\labelwidth - 0.25, \ypos) {Rendering (RTF/ODS)};
\fill[catD, rounded corners=1.5pt]
    (\labelwidth, \ypos - \barheight/2) rectangle (\labelwidth + \bw, \ypos + \barheight/2);
\draw[catD!70!black, line width=0.6pt, rounded corners=1.5pt]
    (\labelwidth, \ypos - \barheight/2) rectangle (\labelwidth + \bw, \ypos + \barheight/2);
\node[anchor=east, font=\sffamily\footnotesize\bfseries, white]
    at (\labelwidth + \bw - 0.15, \ypos) {80};
\node[anchor=west, font=\sffamily\footnotesize\bfseries, catD!80!black]
    at (\labelwidth + \bw + 0.2, \ypos) {14\%};

\pgfmathsetmacro{\ypos}{-5*\rowspacing}
\pgfmathsetmacro{\bw}{60/\maxval*\maxwidth}
\node[anchor=east, font=\sffamily\small] at (\labelwidth - 0.25, \ypos) {Data Preparation};
\fill[catE, rounded corners=1.5pt]
    (\labelwidth, \ypos - \barheight/2) rectangle (\labelwidth + \bw, \ypos + \barheight/2);
\draw[catE!70!black, line width=0.6pt, rounded corners=1.5pt]
    (\labelwidth, \ypos - \barheight/2) rectangle (\labelwidth + \bw, \ypos + \barheight/2);
\node[anchor=west, font=\sffamily\footnotesize\bfseries, catE!80!black]
    at (\labelwidth + \bw + 0.2, \ypos) {60~~(11\%)};

\pgfmathsetmacro{\ypos}{-6*\rowspacing}
\pgfmathsetmacro{\bw}{48/\maxval*\maxwidth}
\node[anchor=east, font=\sffamily\small] at (\labelwidth - 0.25, \ypos) {Orchestration \& Control};
\fill[catF, rounded corners=1.5pt]
    (\labelwidth, \ypos - \barheight/2) rectangle (\labelwidth + \bw, \ypos + \barheight/2);
\draw[catF!70!black, line width=0.6pt, rounded corners=1.5pt]
    (\labelwidth, \ypos - \barheight/2) rectangle (\labelwidth + \bw, \ypos + \barheight/2);
\node[anchor=west, font=\sffamily\footnotesize\bfseries, catF!80!black]
    at (\labelwidth + \bw + 0.2, \ypos) {48~~(9\%)};

\node[rectangle, rounded corners=2.5pt, draw=gray!40, fill=gray!6,
      inner sep=5pt, font=\sffamily\small\bfseries, text=gray!60!black,
      anchor=east] at (\labelwidth + \maxwidth + 2.0, -7*\rowspacing - 0.6)
      {\textit{N}\,=\,558 components};

\end{tikzpicture}}
    \caption{Component taxonomy of the legacy 558-macro library, classified by functional role. Formatting and display macros constitute the largest category (27\%), reflecting the high proportion of code dedicated to output appearance rather than statistical computation.}
    \label{fig:taxonomy}
\end{figure}
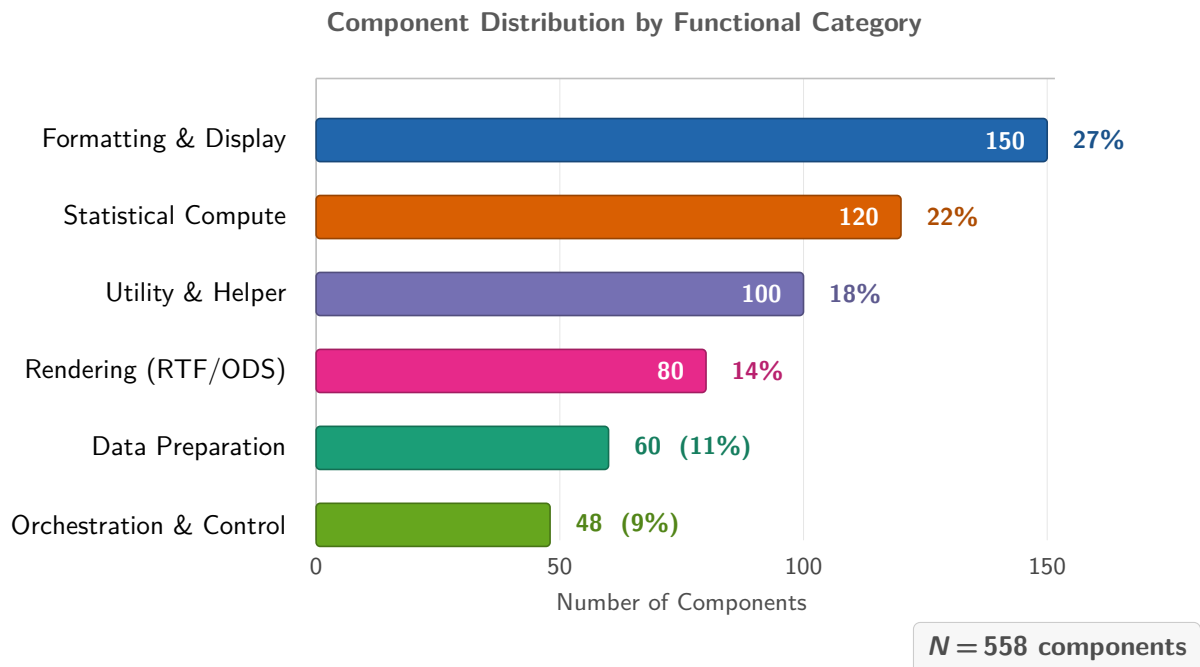

The four activities produce a structured profile: a classified component inventory (\tabref{tab:metrics}, legacy column), an annotated dependency graph, a complexity distribution with outliers flagged, and a coverage matrix. \figref{fig:taxonomy} presents the taxonomy breakdown for the case-study library. The methodology requires no proprietary tools and is applicable to any pharmaceutical SAS macro library.

\subsection{Architecture Design}\label{sec:arch}

The framework architecture addresses the structural deficiencies identified in \secref{sec:eval} through a layered, metadata-driven pipeline with explicit contracts at every phase boundary. This section presents the design principles, pipeline structure, IR, configuration model, AI-cooperative interfaces, regulatory compliance architecture, and the non-destructive adoption pattern that allows the architecture to be deployed without modifying legacy SAS source.

\subsubsection{Design Principles}

Five principles guide the architecture:

\textbf{Separation of concerns.} Each macro performs exactly one function and declares its inputs and outputs through a standard interface, eliminating the monolithic macros identified in the legacy evaluation.

\textbf{Metadata-driven configuration.} YAML configuration files replace hardcoded parameters. Report definitions, population filters, treatment arm labels, and pipeline sequences are specified declaratively. Modifying a report requires editing YAML, not SAS code~[35].

\textbf{Format-agnostic intermediate representation.} An IR sits between computation and rendering. The compute layer produces results in a standard schema; the render layer reads a typed cell grid and emits RTF, PDF, HTML, or JSON. Adding a new output format requires only a new render engine.

\textbf{AI-cooperative interfaces.} All pipeline artifacts --- configurations, IRs, execution manifests --- are available in structured formats (YAML, JSON), enabling LLMs and AI agents to read results, generate configurations, and participate in quality control without parsing proprietary formats.

\textbf{Regulatory compliance by design.} Audit trails, execution manifests, input/output hashing, and electronic signature support are architectural requirements. Every pipeline step records provenance metadata sufficient to reconstruct how a given output was produced.

\subsubsection{Five-Phase Pipeline}

\begin{figure}[H]
    \centering
    \resizebox{\textwidth}{!}{
\begin{tikzpicture}[
    font=\sffamily\small,
    >=Stealth,
    node distance=3mm and 5mm,
    phase/.style={
        rectangle, rounded corners=3pt, draw=#1!70!black, line width=0.7pt,
        fill=#1!12, minimum width=24mm, minimum height=14mm, align=center,
        text width=21mm, inner sep=3pt
    },
    phase/.default=blue,
    config/.style={
        rectangle, rounded corners=2pt, draw=orange!65!black, line width=0.6pt,
        fill=orange!8, minimum width=20mm, minimum height=9mm, align=center,
        text width=18mm, inner sep=2pt, font=\sffamily\footnotesize
    },
    contract/.style={
        font=\sffamily\footnotesize, text=gray!60!black, align=center, text width=17mm
    },
    flow/.style={->, line width=0.9pt, gray!65!black},
    configflow/.style={->, dashed, line width=0.6pt, orange!55!black},
    crosscut/.style={
        rectangle, rounded corners=2pt, draw=gray!45, line width=0.6pt,
        fill=gray!5, minimum height=7.5mm, align=center, inner sep=3pt,
        font=\sffamily\footnotesize\itshape
    },
]

\definecolor{phA}{HTML}{4393C3}   
\definecolor{phB}{HTML}{5AAE61}   
\definecolor{phC}{HTML}{F4A582}   
\definecolor{phD}{HTML}{2166AC}   
\definecolor{phE}{HTML}{B2182B}   

\node[phase=phA] (p1) {\footnotesize\textbf{Phase 1}\\\footnotesize ADaM\\\footnotesize Derivation};
\node[phase=phB, right=8mm of p1] (p2) {\footnotesize\textbf{Phase 2}\\\footnotesize Data\\\footnotesize Preparation};
\node[phase=phC, right=8mm of p2] (p3) {\footnotesize\textbf{Phase 3}\\\footnotesize Compute};
\node[phase=phD, right=8mm of p3, line width=1.0pt, fill=phD!18] (p4) {\footnotesize\textbf{Phase 4}\\\footnotesize IR\\\footnotesize Transform};
\node[phase=phE, right=8mm of p4] (p5) {\footnotesize\textbf{Phase 5}\\\footnotesize Render};

\draw[flow] (p1) -- (p2);
\draw[flow] (p2) -- (p3);
\draw[flow] (p3) -- (p4);
\draw[flow] (p4) -- (p5);

\node[contract, above=1mm] at ($(p1.east)!0.5!(p2.west)$) {ADaM\\datasets};
\node[contract, above=1mm] at ($(p2.east)!0.5!(p3.west)$) {Analysis-\\ready data};
\node[contract, above=1mm] at ($(p3.east)!0.5!(p4.west)$) {Standard\\schema};
\node[contract, above=1mm] at ($(p4.east)!0.5!(p5.west)$) {IR cells +\\structure};

\node[config, above=14mm of p1] (cfg1) {Study\\Config};
\node[config, above=14mm of p2] (cfg2) {Report Type\\Registry};
\node[config, above=14mm of p3] (cfg3) {Compute\\Params};
\node[config, above=14mm of p4] (cfg4) {IR Mapping\\Config};
\node[config, above=14mm of p5] (cfg5) {Render\\Config};

\node[font=\sffamily\small\bfseries, orange!60!black, anchor=south]
    at ($(cfg1.north west)!0.5!(cfg5.north east) + (0,3mm)$) {YAML Configuration Layer};

\draw[configflow] (cfg1) -- (p1);
\draw[configflow] (cfg2) -- (p2);
\draw[configflow] (cfg3) -- (p3);
\draw[configflow] (cfg4) -- (p4);
\draw[configflow] (cfg5) -- (p5);

\node[crosscut, minimum width=132mm, below=8mm of p3]
    (xcut) {Audit Trail (\texttt{fw\_audit}) \quad|\quad Error Handling \quad|\quad Execution Manifest \quad|\quad Provenance};

\node[font=\sffamily\footnotesize, anchor=west, text width=14mm, align=left]
    at ($(p5.east) + (4mm, 0)$) {RTF\\PDF\\HTML\\JSON};
\draw[->, line width=0.7pt, gray!55!black] (p5.east) -- ++(3mm, 0);

\node[rectangle, rounded corners=2pt, draw=gray!35, fill=white, inner sep=4pt,
      font=\sffamily\footnotesize, anchor=north west, text width=34mm, align=left]
    at ([yshift=-3mm]xcut.south west) {%
    \textcolor{phD}{\rule{5pt}{5pt}}~Pipeline phase\\
    \textcolor{orange!60!black}{\rule{5pt}{5pt}}~YAML config\\
    \textcolor{gray!50}{\rule{5pt}{5pt}}~Cross-cutting
};

\end{tikzpicture}}
    \caption{Five-phase pipeline architecture with explicit input/output contracts at each layer boundary. YAML configuration (top) feeds into every phase, while the Intermediate Representation between Compute and Render serves as the key decoupling contract. Cross-cutting concerns span all phases.}
    \label{fig:pipeline}
\end{figure}
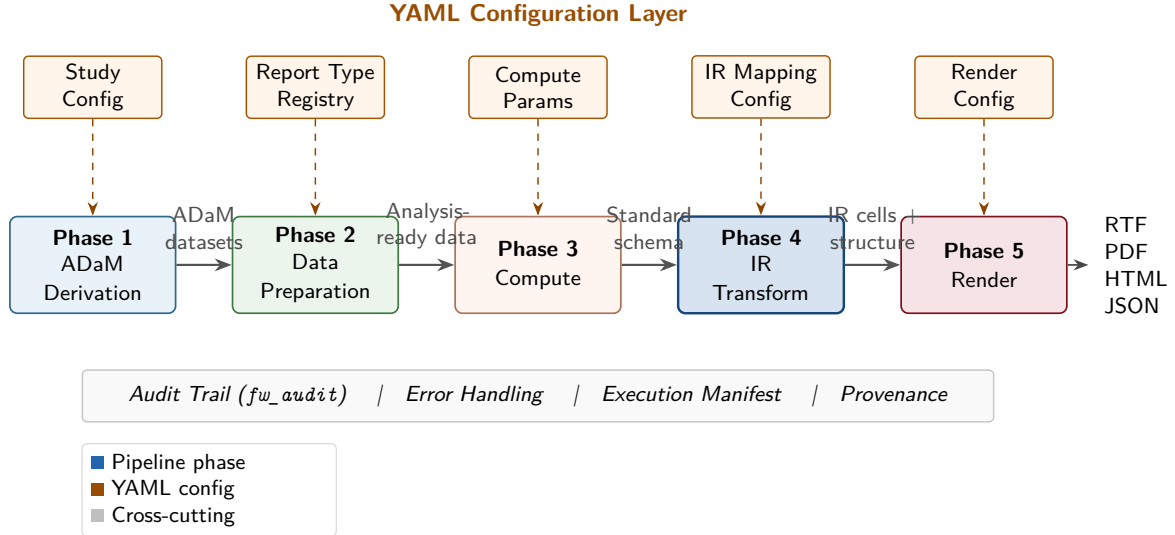

The framework organizes clinical report production into five sequential phases with explicit input and output contracts (\figref{fig:pipeline}).

\textbf{Phase 1: ADaM Derivation.} This optional phase generates ADaM datasets from SDTM source data using YAML-configured derivation rules. Seven macros handle major ADaM structures: ADSL, BDS, OCCDS, and ADTTE, plus generic derivation, safety algorithm, and validation modules.

\textbf{Phase 2: Data Preparation.} Data preparation macros transform ADaM datasets into analysis-ready form: subsetting, derivation, flag creation, ADSL merging, row completion, treatment sorting, and total column addition. Every macro follows a standard interface (\code{input\_ds}, \code{output\_ds}, operation-specific parameters, \code{execution\_id}, \code{debug}). The pipeline sequence for a given report type is specified in YAML.

\textbf{Phase 3: Compute.} Compute macros execute statistical analyses on prepared data, covering descriptive statistics, frequency counts, incidence rates, Kaplan-Meier estimation, log-rank tests, chi-square tests, shift tables, change from baseline, ANOVA, regression, mixed models, and subgroup analyses. Every compute macro produces output in a standard schema containing grouping variables, treatment identifiers, statistic names, numeric values, formatted strings, and method identifiers. This standard schema is the critical contract enabling downstream phases to operate uniformly.

\textbf{Phase 4: IR Transform.} Three components convert compute output into the IR (described in \secref{sec:arch:ir}): \code{cl\_to\_ir} performs the transformation, \code{cl\_ir\_reconcile} verifies numeric integrity by tracing cells back to source statistics within configurable tolerance (default $1\times10^{-10}$), and \code{fw\_validate\_ir} checks structural constraints.

\textbf{Phase 5: Render.} Render engines read IR datasets and produce final output. SAS engines generate RTF for regulatory submission; Python engines generate JSON for AI consumption, HTML for interactive review, and PDF for archival. Render engines contain no statistical logic --- they are pure format translators.

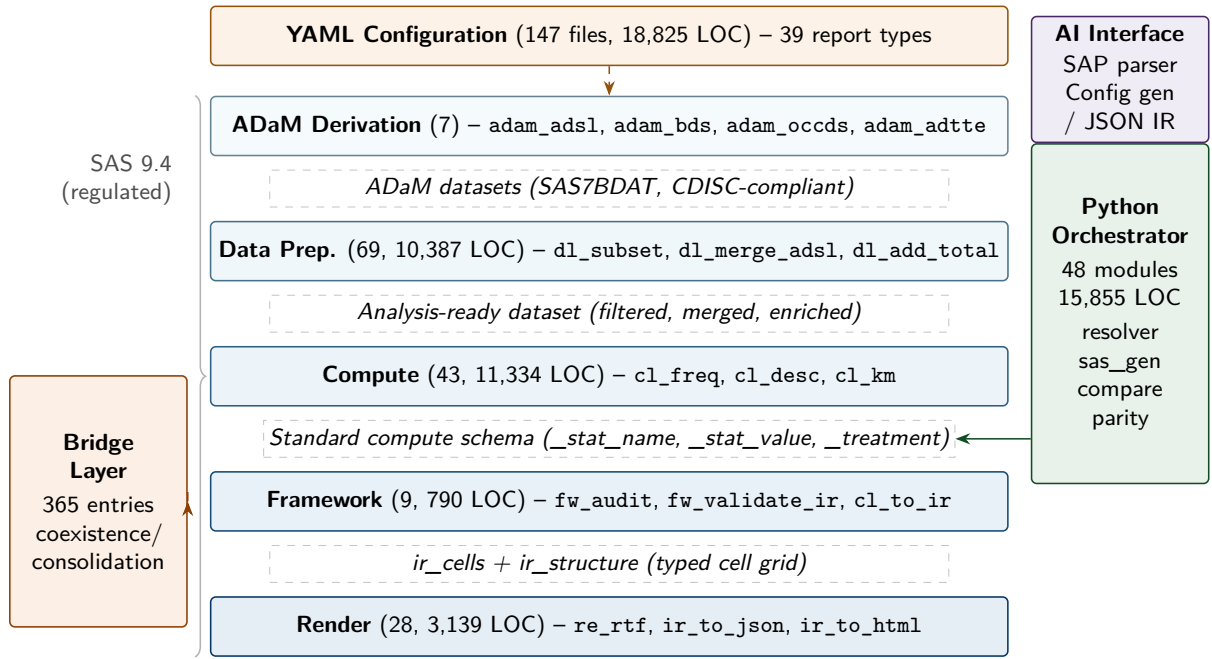
\begin{figure}[htbp]
    \centering
    \resizebox{\textwidth}{!}{
\begin{tikzpicture}[
    font=\sffamily\small,
    >=Stealth,
    layer/.style={
        rectangle, rounded corners=2pt, draw=#1!65!black, line width=0.6pt,
        fill=#1!10, minimum height=8mm, align=center, inner sep=3pt,
        font=\sffamily\footnotesize
    },
    layer/.default=blue,
    contract/.style={
        rectangle, draw=gray!45, dashed, line width=0.4pt,
        fill=white, minimum height=5mm, align=center, inner sep=2pt,
        font=\sffamily\footnotesize\itshape
    },
    sidebar/.style={
        rectangle, rounded corners=2pt, draw=#1!65!black, line width=0.6pt,
        fill=#1!10, align=center, inner sep=3pt
    },
    sidebar/.default=teal,
]

\definecolor{layR}{HTML}{08519C}   
\definecolor{layF}{HTML}{2171B5}   
\definecolor{layC}{HTML}{4292C6}   
\definecolor{layD}{HTML}{6BAED6}   
\definecolor{layA}{HTML}{9ECAE1}   
\definecolor{layY}{HTML}{E08214}   
\definecolor{layBr}{HTML}{D95F02}  
\definecolor{layPy}{HTML}{1B7837}  
\definecolor{layAI}{HTML}{6A3D9A}  

\def\layerwidth{108mm}
\def\layerheight{8mm}
\def\sidebw{24mm}
\def\cgap{1.8mm}

\node[layer=layR, minimum width=\layerwidth, minimum height=\layerheight]
    (render) at (0, 0) {\textbf{Render} (28, 3{,}139 LOC) -- \texttt{re\_rtf}, \texttt{ir\_to\_json}, \texttt{ir\_to\_html}};

\node[contract, minimum width=92mm, above=\cgap of render]
    (c5) {ir\_cells + ir\_structure (typed cell grid)};

\node[layer=layF, minimum width=\layerwidth, minimum height=\layerheight, above=\cgap of c5]
    (fw) {\textbf{Framework} (9, 790 LOC) -- \texttt{fw\_audit}, \texttt{fw\_validate\_ir}, \texttt{cl\_to\_ir}};

\node[contract, minimum width=92mm, above=\cgap of fw]
    (c4) {Standard compute schema (\_stat\_name, \_stat\_value, \_treatment)};

\node[layer=layC, minimum width=\layerwidth, minimum height=\layerheight, above=\cgap of c4]
    (compute) {\textbf{Compute} (43, 11{,}334 LOC) -- \texttt{cl\_freq}, \texttt{cl\_desc}, \texttt{cl\_km}};

\node[contract, minimum width=92mm, above=\cgap of compute]
    (c3) {Analysis-ready dataset (filtered, merged, enriched)};

\node[layer=layD, minimum width=\layerwidth, minimum height=\layerheight, above=\cgap of c3]
    (dataprep) {\textbf{Data Prep.} (69, 10{,}387 LOC) -- \texttt{dl\_subset}, \texttt{dl\_merge\_adsl}, \texttt{dl\_add\_total}};

\node[contract, minimum width=92mm, above=\cgap of dataprep]
    (c2) {ADaM datasets (SAS7BDAT, CDISC-compliant)};

\node[layer=layA, minimum width=\layerwidth, minimum height=\layerheight, above=\cgap of c2]
    (adam) {\textbf{ADaM Derivation} (7) -- \texttt{adam\_adsl}, \texttt{adam\_bds}, \texttt{adam\_occds}, \texttt{adam\_adtte}};

\node[layer=layY, minimum width=\layerwidth, minimum height=\layerheight, above=4mm of adam]
    (yaml) {\textbf{YAML Configuration} (147 files, 18{,}825 LOC) -- 39 report types};

\node[sidebar=layBr, minimum width=\sidebw, minimum height=34mm,
      text width=20mm, align=center, anchor=east,
      font=\sffamily\footnotesize]
    (bridge) at ([xshift=-3mm]render.south west |- fw.center) {%
    \textbf{Bridge}\\
    \textbf{Layer}\\[2pt]
    365 entries\\
    coexistence/\\
    consolidation};

\node[sidebar=layPy, minimum width=\sidebw, minimum height=46mm,
      text width=20mm, align=center, anchor=west,
      font=\sffamily\footnotesize]
    (python) at ([xshift=3mm]render.south east |- c3.center) {%
    \textbf{Python}\\
    \textbf{Orchestrator}\\[2pt]
    48 modules\\
    15{,}855 LOC\\[2pt]
    resolver\\
    sas\_gen\\
    compare\\
    parity};

\node[sidebar=layAI, minimum width=\sidebw, minimum height=13mm,
      text width=20mm, align=center, anchor=south west,
      font=\sffamily\footnotesize]
    (ai) at (python.north west) {%
    \textbf{AI Interface}\\[1pt]
    SAP parser\\
    Config gen / JSON IR};

\draw[->, dashed, layY!65!black, line width=0.6pt] (yaml.south) -- (adam.north);
\draw[->, line width=0.6pt, layPy!65!black] (python.west |- c4.east) -- (c4.east);
\draw[->, line width=0.6pt, layBr!65!black] (bridge.east) -- ([xshift=-3mm]fw.west |- bridge.east);

\draw[decorate, decoration={brace, amplitude=4pt, mirror}, gray!60, line width=0.6pt]
    ([xshift=-2mm]render.south west) -- ([xshift=-2mm]adam.north west)
    node[pos=0.85, left=5pt, font=\sffamily\footnotesize, align=right, text=gray!60!black] {SAS 9.4\\(regulated)};

\end{tikzpicture}}
    \caption{Layered architecture showing six SAS layers (ADaM Derivation, Data Preparation, Compute, Framework, Bridge Compatibility, Render) with explicit contracts at each boundary, plus the Python orchestration layer and AI interfaces. The bridge layer provides a 365-entry facade over the legacy library.}
    \label{fig:contracts}
\end{figure}

Each phase boundary enforces an explicit contract: the output schema of phase $N$ is the required input of phase $N{+}1$, enabling independent testing, replacement, or extension of any phase (\figref{fig:contracts}).

\subsubsection{Intermediate Representation}\label{sec:arch:ir}

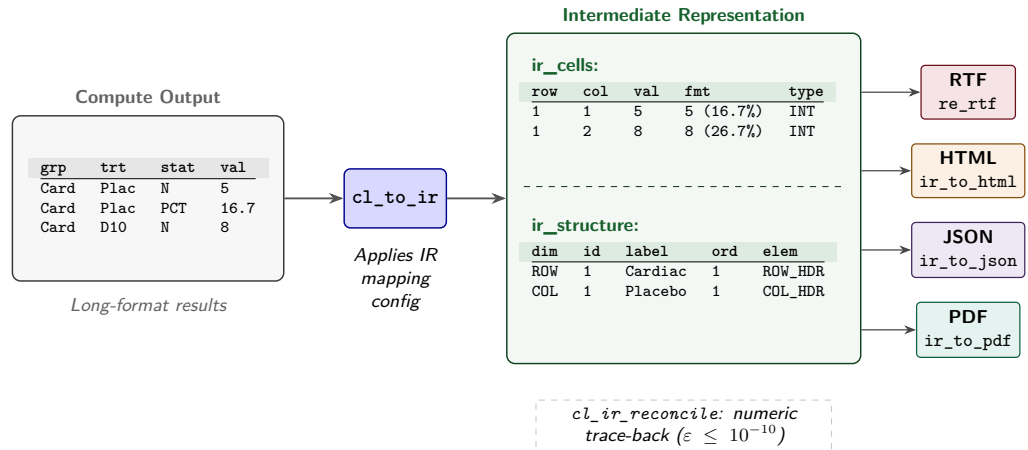
\begin{figure}[H]
    \centering
    \resizebox{\textwidth}{!}{
\begin{tikzpicture}[
    font=\sffamily\small,
    >=Stealth,
    node distance=3mm and 5mm,
    zone/.style={
        rectangle, rounded corners=4pt, draw=#1!55!black, line width=0.7pt,
        fill=#1!5, inner sep=5pt, align=center
    },
    zone/.default=blue,
    process/.style={
        rectangle, rounded corners=3pt, draw=blue!55!black, line width=0.7pt,
        fill=blue!15, minimum width=17mm, minimum height=10mm, align=center,
        font=\sffamily\small\bfseries
    },
    renderer/.style={
        rectangle, rounded corners=2pt, draw=#1!60!black, line width=0.6pt,
        fill=#1!12, minimum width=16mm, minimum height=8mm, align=center,
        font=\sffamily\footnotesize
    },
    renderer/.default=green,
]

\definecolor{irGreen}{HTML}{1B7837}   
\definecolor{compGray}{HTML}{636363}  
\definecolor{rendRTF}{HTML}{B2182B}   
\definecolor{rendHTML}{HTML}{E08214}  
\definecolor{rendJSON}{HTML}{6A3D9A}  
\definecolor{rendPDF}{HTML}{1B9E77}   

\node[zone=compGray, minimum width=46mm, minimum height=28mm,
      label={[font=\sffamily\footnotesize\bfseries, compGray]above:Compute Output}] (z1) at (0,0) {};

\node[font=\ttfamily\scriptsize, align=left] at (z1.center) {%
\begin{tabular}{@{}llll@{}}
\rowcolor{gray!20}
\scriptsize\bfseries grp & \scriptsize\bfseries trt & \scriptsize\bfseries stat & \scriptsize\bfseries val \\
\hline
Card & Plac & N & 5 \\
Card & Plac & PCT & 16.7 \\
Card & D10 & N & 8 \\
\end{tabular}%
};
\node[font=\sffamily\footnotesize\itshape, compGray, anchor=north]
    at ([yshift=-1.5mm]z1.south) {Long-format results};

\node[process, right=10mm of z1] (transform) {\texttt{cl\_to\_ir}};
\node[font=\sffamily\footnotesize\itshape, text width=18mm, align=center, below=2mm of transform]
    {Applies IR\\mapping config};

\draw[->, line width=0.9pt, gray!65!black] (z1.east) -- (transform.west);

\node[zone=irGreen, minimum width=60mm, minimum height=56mm, right=10mm of transform,
      label={[font=\sffamily\footnotesize\bfseries, irGreen!80!black]above:Intermediate Representation}] (z2) {};

\node[font=\sffamily\footnotesize\bfseries, irGreen!80!black, anchor=north west]
    at ([xshift=3mm, yshift=-3mm]z2.north west) {ir\_cells:};
\node[font=\ttfamily\scriptsize, align=left, anchor=north west]
    at ([xshift=3mm, yshift=-7mm]z2.north west) {%
\begin{tabular}{@{}lllll@{}}
\rowcolor{irGreen!15}
\scriptsize\bfseries row & \scriptsize\bfseries col & \scriptsize\bfseries val & \scriptsize\bfseries fmt & \scriptsize\bfseries type \\
\hline
1 & 1 & 5 & 5 (16.7\%) & INT \\
1 & 2 & 8 & 8 (26.7\%) & INT \\
\end{tabular}%
};

\draw[irGreen!40!black, line width=0.4pt, dashed]
    ([xshift=3mm, yshift=2mm]z2.west) -- ([xshift=-3mm, yshift=2mm]z2.east);

\node[font=\sffamily\footnotesize\bfseries, irGreen!80!black, anchor=north west]
    at ([xshift=3mm, yshift=-2mm]z2.west |- z2.center) {ir\_structure:};
\node[font=\ttfamily\scriptsize, align=left, anchor=north west]
    at ([xshift=3mm, yshift=-6mm]z2.west |- z2.center) {%
\begin{tabular}{@{}lllll@{}}
\rowcolor{irGreen!15}
\scriptsize\bfseries dim & \scriptsize\bfseries id & \scriptsize\bfseries label & \scriptsize\bfseries ord & \scriptsize\bfseries elem \\
\hline
ROW & 1 & Cardiac & 1 & ROW\_HDR \\
COL & 1 & Placebo & 1 & COL\_HDR \\
\end{tabular}%
};

\draw[->, line width=0.9pt, gray!65!black] (transform.east) -- (z2.west);

\node[renderer=rendRTF,    right=10mm of z2, yshift=18mm] (r1) {\textbf{RTF}\\\footnotesize\texttt{re\_rtf}};
\node[renderer=rendHTML,   below=4mm of r1] (r2) {\textbf{HTML}\\\footnotesize\texttt{ir\_to\_html}};
\node[renderer=rendJSON,   below=4mm of r2] (r3) {\textbf{JSON}\\\footnotesize\texttt{ir\_to\_json}};
\node[renderer=rendPDF,    below=4mm of r3] (r4) {\textbf{PDF}\\\footnotesize\texttt{ir\_to\_pdf}};

\draw[->, line width=0.6pt, gray!55!black] (z2.east |- r1) -- (r1.west);
\draw[->, line width=0.6pt, gray!55!black] (z2.east |- r2) -- (r2.west);
\draw[->, line width=0.6pt, gray!55!black] (z2.east |- r3) -- (r3.west);
\draw[->, line width=0.6pt, gray!55!black] (z2.east |- r4) -- (r4.west);

\node[rectangle, draw=gray!45, dashed, fill=white, inner sep=3pt,
      font=\sffamily\footnotesize\itshape, text width=48mm, align=center,
      below=6mm of z2]
    (reconc) {\texttt{cl\_ir\_reconcile}: numeric trace-back ($\varepsilon \leq 10^{-10}$)};

\node[font=\sffamily\footnotesize, align=center, anchor=north,
      gray!65!black] at ([yshift=-5mm]reconc.south -| transform) {%
    \textbf{cell\_type vocabulary:}~
    INTEGER~\textbullet~DECIMAL~\textbullet~PVALUE~\textbullet~PERCENTAGE~\textbullet~TEXT~\textbullet~HEADER~\textbullet~LABEL~\textbullet~FOOTNOTE~\textbullet~EMPTY};

\end{tikzpicture}}
    \caption{Intermediate Representation (IR) schema showing the decoupling of statistical computation from rendering. The \texttt{cl\_to\_ir} macro transforms long-format compute output (left) into a grid-based cell model consisting of \texttt{ir\_cells} and \texttt{ir\_structure} datasets (centre). The format-agnostic IR feeds multiple render engines (right). Numeric integrity is verified by \texttt{cl\_ir\_reconcile}.}
    \label{fig:ir}
\end{figure}

The IR is the central architectural innovation, representing a clinical report as two complementary datasets (\figref{fig:ir}).

\textbf{ir\_cells} contains one row per cell in the report grid: \code{row\_id}, \code{col\_id}, raw numeric \code{cell\_value}, formatted display string \code{cell\_formatted}, and \code{cell\_type} from a controlled vocabulary (INTEGER, DECIMAL, PVALUE, PERCENTAGE, TEXT, HEADER, LABEL, FOOTNOTE, EMPTY). Numeric types carry both raw values and formatted strings; non-numeric types carry only display content. This controlled vocabulary is the key enabler of AI-driven pharmacoinformatics (\secref{sec:ai}): by semantically classifying each cell, the vocabulary allows downstream consumers --- including large language models --- to distinguish a p-value from a percentage, a column header from a data row, and a statistical result from a footnote, without heuristic parsing of rendered output. Every cell also carries \code{report\_id}, \code{execution\_id}, and \code{sort\_order}.

\textbf{ir\_structure} defines the row and column dimensions: \code{dimension} (ROW or COL), \code{dim\_id} linking to ir\_cells, display label, sort order, indent level, text alignment, span count, and \code{element\_type} (COLUMN\_HEADER, ROW\_HEADER, DATA\_ROW, TOTAL\_ROW, SEPARATOR, SPANNING\_HEADER).

Four structural rules govern IR validity: \textbf{completeness} (every cell pair has corresponding structure definitions), \textbf{contiguity} (identifiers start at 1 with no gaps), \textbf{uniqueness} (each grid position appears once), and \textbf{consistency} (\code{report\_id} and \code{execution\_id} match across datasets).

This design enables cell-level QC on structured data rather than rendered text, format-agnostic output from a single IR, and reconciliation against source compute output.

\subsubsection{YAML Registry and Configuration}

The configuration model has three tiers. The \textbf{report type registry} contains one YAML file per report type, defining pipeline steps, parameters, IR mapping rules, and render configuration. The registry currently defines 39 report types. \textbf{Study-level configuration} provides per-study overrides: populations, treatment labels, dataset locations. \textbf{Parameter resolution} occurs through a Python bridge (\code{yaml\_to\_sas}) that reads YAML, resolves cross-references, and emits SAS macro variables, handling type conversion and the SAS 32-character name limit.

\subsubsection{AI-Cooperative Interfaces}

The architecture provides three AI integration points. \textbf{JSON export} (\code{ir\_to\_json}) exports IR datasets as structured JSON with typed columns and cell arrays, enabling LLMs to reason about clinical results as structured data. \textbf{SAP document parser} (\code{sap\_parser}) extracts table specifications from SAP documents. \textbf{Configuration generator} (\code{config\_generator}) produces draft YAML configurations from parsed SAP output, with a \code{matcher} module mapping specifications to registry entries. Generated configurations require human review --- the framework inserts an explicit review gate.

The design choice is that AI agents interact with IR and YAML layers, not raw SAS output. Structured data enables structured reasoning: an LLM can identify that a p-value cell contains 0.0312 and the method is chi-square, without parsing RTF layout.

\subsubsection{Regulatory Compliance Architecture}

\textbf{Audit trail.} The \code{fw\_audit} macro records every pipeline step with timestamp, user identifier, event type, step name, status, layer, and comments. Records accumulate in a persistent SAS dataset.

\textbf{Execution manifest.} Each pipeline run records input datasets with file hashes, resolved parameters, macro versions, SAS session metadata, and output hashes --- sufficient to answer ``what exact inputs and parameters produced this output?''

\textbf{Electronic signature support.} Audit trail and manifest structures are designed to support 21 CFR Part 11 requirements~[36], including user authentication linkage and tamper-evident storage.

\textbf{Change control.} YAML report definitions in version control produce reviewable, diff-able commits readable by both programmers and non-programmer reviewers.

The combination of audit trail, execution manifest, and version-controlled configuration creates a traceability chain from submission output through every transformation step to source ADaM datasets and SAP specification~[37].

\subsubsection{Non-Destructive Adoption Pattern}\label{sec:arch:nda}

The bridge map is the single artifact that distinguishes the two deployment modes. The Native Driver does not need to know which mode applies to any given entry: it simply executes the entry's \code{native\_target}. The deployment surface that an organization manipulates over time is therefore the metadata, not the SAS source. This indirection has three consequences. First, the regulatory standing of the legacy macros is preserved by default, because their files are read but never written by the framework. Second, the rate of consolidation is a business decision rather than an architectural commitment --- an organization may consolidate aggressively, conservatively, or not at all, using the same framework. Third, AI readiness is decoupled from consolidation: the IR and JSON export are produced by the metadata layer regardless of whether the underlying computation originated in a legacy macro running in coexistence mode or in a modern core macro running in consolidation mode. A pharmaceutical organization can therefore integrate AI workflows on day one without committing to any source-level modernization.

\textbf{Polyglot integration.} The non-destructive adoption pattern is what makes polyglot integration practical in a regulated setting. SAS retains exclusive responsibility for regulated computation, including in coexistence mode where the original macros run unchanged within their existing validation envelope. Python operates outside the regulated boundary, providing pipeline orchestration, IR-to-JSON serialization, the SAP parser, and the parity validation harness. R consumes the JSON IR through standard parsers (e.g., \code{jsonlite}) for exploratory analytics, custom visualizations, and statistical cross-checks, without re-implementing the validated SAS computation. Large language models receive the same JSON IR and operate on its typed cell structure, eliminating the brittle RTF parsing that has prevented LLM integration in legacy workflows. Each language addresses a single concern, the IR is the single integration surface, and the legacy library remains the authoritative source of regulated computational logic. This division of labor is consistent with the recommendations of GAMP~5~[40] for computerized systems in regulated industries, which advocates strict separation between validated computational components and unvalidated supporting infrastructure, and with broader software-engineering guidance on legacy system migration in safety-critical and regulated domains~[29].

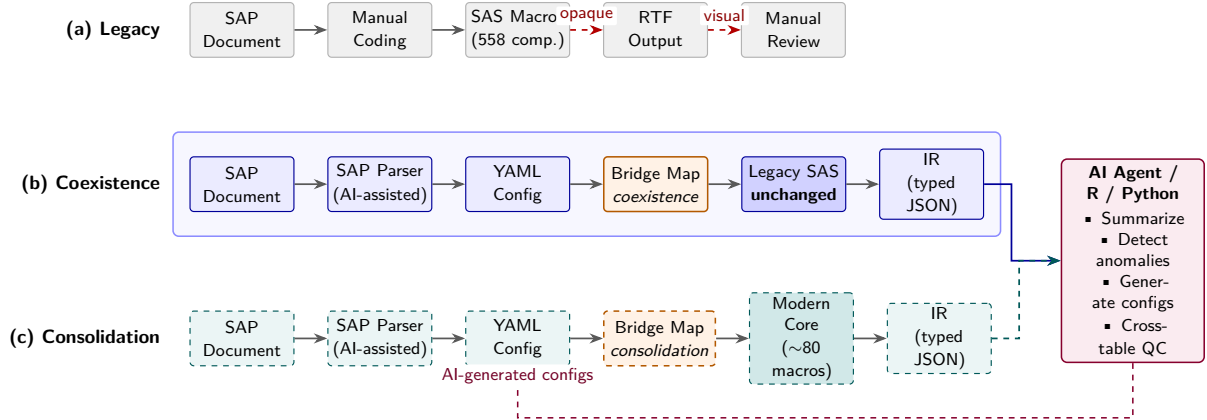
\begin{figure}[htbp]
    \centering
    \resizebox{\textwidth}{!}{

\begin{tikzpicture}[
    font=\sffamily\small,
    >=Stealth,
    node distance=4mm and 6mm,
    legacybox/.style={
        rectangle, rounded corners=2pt, draw=gray!60, line width=0.6pt,
        fill=gray!12,
        minimum width=19mm, minimum height=10mm, align=center,
        text width=17mm, inner sep=2pt
    },
    coexbox/.style={
        rectangle, rounded corners=2pt, draw=blue!60!black, line width=0.6pt,
        fill=blue!8, minimum width=19mm, minimum height=10mm, align=center,
        text width=17mm, inner sep=2pt
    },
    consbox/.style={
        rectangle, rounded corners=2pt, draw=teal!70!black, line width=0.6pt,
        dashed, fill=teal!8, minimum width=19mm, minimum height=10mm,
        align=center, text width=17mm, inner sep=2pt
    },
    bridgebox/.style={
        rectangle, rounded corners=2pt, draw=orange!70!black, line width=0.8pt,
        fill=orange!10,
        minimum width=19mm, minimum height=10mm, align=center,
        text width=17mm, inner sep=2pt
    },
    aibox/.style={
        rectangle, rounded corners=3pt, draw=purple!70!black, line width=0.8pt,
        fill=purple!8, minimum width=26mm, minimum height=24mm, align=center,
        text width=24mm, inner sep=3pt
    },
    rowlabel/.style={font=\sffamily\bfseries\small, anchor=east},
    flow/.style={->, line width=0.9pt, gray!70!black},
    barrier/.style={->, line width=0.9pt, red!70!black, dashed},
]

\node[legacybox] (l1) {\footnotesize SAP\\Document};
\node[legacybox, right=of l1] (l2) {\footnotesize Manual\\Coding};
\node[legacybox, right=of l2] (l3) {\footnotesize SAS Macros\\(558 comp.)};
\node[legacybox, right=of l3] (l4) {\footnotesize RTF\\Output};
\node[legacybox, right=of l4] (l5) {\footnotesize Manual\\Review};

\draw[flow] (l1) -- (l2);
\draw[flow] (l2) -- (l3);
\draw[barrier] (l3) -- (l4)
    node[midway, above=2pt, font=\sffamily\footnotesize, text=red!55!black,
         fill=white, inner sep=2pt, rounded corners=1pt] {opaque};
\draw[barrier] (l4) -- (l5)
    node[midway, above=2pt, font=\sffamily\footnotesize, text=red!55!black,
         fill=white, inner sep=2pt, rounded corners=1pt] {visual};

\node[rowlabel] at ([xshift=-4mm]l1.west) {(a) Legacy};

\node[coexbox, below=18mm of l1] (c1) {\footnotesize SAP\\Document};
\node[coexbox, right=of c1] (c2) {\footnotesize SAP Parser\\\footnotesize (AI-assisted)};
\node[coexbox, right=of c2] (c3) {\footnotesize YAML\\Config};
\node[bridgebox, right=of c3] (c4) {\footnotesize Bridge Map\\\footnotesize\textit{coexistence}};
\node[coexbox, right=of c4, fill=blue!18] (c5) {\footnotesize Legacy SAS\\\footnotesize\textbf{unchanged}};
\node[coexbox, right=of c5] (c6) {\footnotesize IR\\\footnotesize (typed JSON)};

\draw[flow] (c1) -- (c2);
\draw[flow] (c2) -- (c3);
\draw[flow] (c3) -- (c4);
\draw[flow] (c4) -- (c5);
\draw[flow] (c5) -- (c6);

\node[rowlabel] at ([xshift=-4mm]c1.west) {(b) Coexistence};

\begin{pgfonlayer}{background}
  \node[rectangle, draw=blue!45, line width=0.8pt, rounded corners=3pt,
        fill=blue!3, fit=(c1)(c6),
        inner xsep=3mm, inner ysep=3mm] (rowbg2) {};
\end{pgfonlayer}

\node[consbox, below=18mm of c1] (m1) {\footnotesize SAP\\Document};
\node[consbox, right=of m1] (m2) {\footnotesize SAP Parser\\\footnotesize (AI-assisted)};
\node[consbox, right=of m2] (m3) {\footnotesize YAML\\Config};
\node[bridgebox, right=of m3, dashed, text width=19mm] (m4) {\footnotesize Bridge Map\\\footnotesize\textit{consolidation}};
\node[consbox, right=of m4, fill=teal!18] (m5) {\footnotesize Modern Core\\\footnotesize ($\sim$80 macros)};
\node[consbox, right=of m5] (m6) {\footnotesize IR\\\footnotesize (typed JSON)};

\draw[flow] (m1) -- (m2);
\draw[flow] (m2) -- (m3);
\draw[flow] (m3) -- (m4);
\draw[flow] (m4) -- (m5);
\draw[flow] (m5) -- (m6);

\node[rowlabel] at ([xshift=-4mm]m1.west) {(c) Consolidation};

\node[aibox, right=14mm of c6, yshift=-14mm] (ai) {%
\footnotesize\textbf{AI Agent / R / Python}\\[2pt]
\footnotesize
\textbullet\ Summarize\\
\textbullet\ Detect anomalies\\
\textbullet\ Generate configs\\
\textbullet\ Cross-table QC};

\draw[flow, blue!60!black] (c6.east) -- ++(5mm,0) |- (ai.west);
\draw[flow, teal!70!black, dashed] (m6.east) -- ++(5mm,0) |- (ai.west);

\draw[->, line width=0.8pt, purple!70!black, dashed]
    (ai.south) -- ++(0,-10mm) -| (m3.south)
    node[pos=0.75, above=2pt, font=\sffamily\footnotesize, text=purple!60!black,
         fill=white, inner sep=2pt, rounded corners=1pt] {AI-generated configs};

\end{tikzpicture}}
    \caption{AI integration pathways under the non-destructive adoption pattern... Left column: legacy workflow producing RTF that is opaque to AI agents. Centre column: coexistence-mode adoption, in which the legacy library is preserved unchanged and the metadata layer (bridge map + IR) re-exposes outputs as JSON consumable by LLMs, R, and Python. Right column: optional consolidation-mode adoption, in which selected legacy macros are replaced by parameterized core macros while the same metadata layer continues to produce JSON. The IR contract is identical in the centre and right columns, demonstrating that AI readiness is decoupled from source-level modernization. Coexistence delivers Day-0 AI readiness with no change to legacy SAS source; consolidation rides on the same metadata layer and is applied incrementally.}
    \label{fig:aiflow}
\end{figure}

\textbf{Implications for adoption risk.} The non-destructive pattern materially changes the adoption-risk profile of clinical reporting modernization. Under conventional rewrite approaches, an organization commits up-front to a multi-year project whose value is realized only on completion. Under the coexistence-first pattern proposed here, AI readiness is a Day-0 deliverable obtained by populating the bridge map and standing up the metadata layer; consolidation, where elected, becomes a continuous improvement activity whose individual increments can be evaluated, validated, and accepted independently. This staged value delivery aligns with risk-based validation frameworks such as GAMP~5~[40] and reduces the financial and regulatory commitment required to begin modernization.

\subsection{Migration and Parity Validation}\label{sec:migration}

Modernizing a legacy library is insufficient unless the new framework produces equivalent outputs. This section presents the parity validation methodology: executing legacy and modernized pipelines on identical input data, comparing outputs at the cell level, and triaging discrepancies through a gate-based workflow.

\subsubsection{Bridge Map}

The central artifact is the \emph{bridge map}: a structured registry establishing one-to-one correspondence between each legacy macro invocation and its modernized counterpart. Each entry contains:

\begin{itemize}
    \item \textbf{legacy\_id}: Unique identifier for the legacy macro call (e.g., \code{ae\_summary\_soc}).
    \item \textbf{native\_target}: The modernized pipeline configuration replacing the legacy call.
    \item \textbf{parameter\_mapping}: Explicit mapping from legacy to modern parameter names, accommodating renames and semantic shifts.
    \item \textbf{defaults}: Values for parameters existing only in the new framework, capturing design decisions implicit in legacy code.
    \item \textbf{preamble\_sas}: Optional SAS code establishing preconditions before the native pipeline call.
    \item \textbf{post\_calls}: Optional follow-up processing steps.
\end{itemize}

The bridge map is defined in machine-readable format (YAML or JSON) for programmatic consumption. It documents every migration decision in a single auditable location and enables automated construction of paired execution plans.

\subsubsection{Parity Harness}

The parity harness is a dual-driver execution engine that orchestrates side-by-side comparison of legacy and modernized outputs.

The \textbf{LegacyDriver} accepts a bridge map entry and constructs a SAS program that invokes the original macro with its original parameters on a specified input dataset. It captures the resulting output files (typically RTF), execution logs, and timing data.

The \textbf{NativeDriver} accepts the same bridge map entry and constructs the corresponding modernized pipeline invocation, translating parameters through the bridge map's \code{parameter\_mapping} and \code{defaults} fields, prepending any \code{preamble\_sas}, and appending any \code{post\_calls}. It captures the same artifact types as the LegacyDriver.

Both drivers operate on identical input data loaded from a common source. The harness reads the bridge map and a study-level configuration file that specifies dataset locations, library paths, and environment settings. For each bridge map entry, it executes both drivers, stores their outputs in a structured directory hierarchy organized by report type, and passes the paired outputs to the comparison engine. This dual-driver architecture isolates comparison logic from execution concerns, allowing batch mode across an entire bridge map or selective execution during iterative debugging.

\subsubsection{Validation Strategy}

The comparison engine applies type-specific validation strategies to paired outputs.

\textbf{Cell-level comparison for tables.} The engine parses both RTF files, extracts each cell's textual content, and performs cell-by-cell comparison. Text cells require exact string match after normalization (whitespace trimming, case folding where appropriate). Numeric cells allow an epsilon tolerance to accommodate rounding differences. This approach detects content-level discrepancies that visual inspection would miss.

\textbf{Structural comparison for listings.} Listings are validated by comparing row count, column count, and content across corresponding cells. Because listings are typically long-form patient-level data, structural checks confirm both pipelines select the same records in the same order.

\textbf{Visual comparison for figures.} Figures (e.g., Kaplan-Meier plots) are compared by file size as a coarse structural check, supplemented by manual spot-checking. Full pixel-level comparison is deferred as an optional extension.

Each comparison produces a verdict: \textbf{PASS} (all cells match within tolerance), \textbf{FAIL} (content divergence), \textbf{ERROR} (execution failure), or \textbf{SKIP} (excluded by explicit annotation). Results are aggregated into a summary matrix providing at-a-glance parity status.

\subsubsection{Gate-Based Verification}

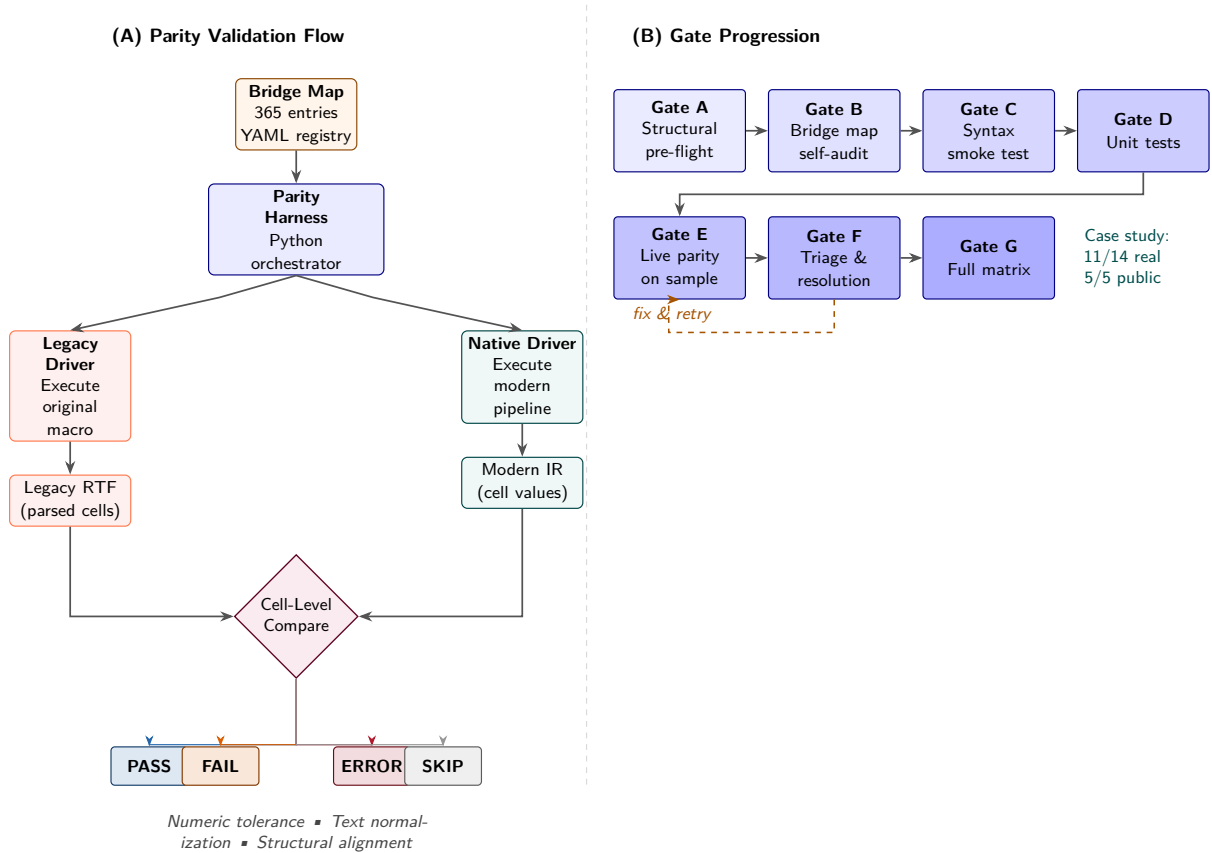
\begin{figure}[H]
    \centering
    \resizebox{\textwidth}{!}{
\begin{tikzpicture}[
    font=\sffamily\small,
    >=Stealth,
    node distance=4mm and 6mm,
    procbox/.style={
        rectangle, rounded corners=3pt, draw=blue!50!black, line width=0.6pt,
        fill=blue!8, minimum width=22mm, minimum height=11mm, align=center,
        text width=20mm, inner sep=3pt
    },
    legbox/.style={
        rectangle, rounded corners=3pt, draw=red!50!orange!70, line width=0.6pt,
        fill=red!6, minimum width=22mm, minimum height=11mm, align=center,
        text width=20mm, inner sep=3pt
    },
    modbox/.style={
        rectangle, rounded corners=3pt, draw=teal!55!black, line width=0.6pt,
        fill=teal!6, minimum width=22mm, minimum height=11mm, align=center,
        text width=20mm, inner sep=3pt
    },
    comparebox/.style={
        diamond, draw=purple!50!black, line width=0.7pt,
        fill=purple!8, minimum width=20mm, minimum height=16mm, align=center,
        inner sep=2pt, font=\sffamily\footnotesize
    },
    resultbox/.style={
        rectangle, rounded corners=2pt, draw=#1!60!black, line width=0.6pt,
        fill=#1!15, minimum width=14mm, minimum height=7mm, align=center,
        font=\sffamily\footnotesize\bfseries
    },
    gate/.style={
        rectangle, rounded corners=2pt, draw=blue!50!black, line width=0.6pt,
        fill=blue!#1, minimum width=24mm, minimum height=15mm, align=center,
        text width=21mm, inner sep=3pt
    },
    flow/.style={->, line width=0.9pt, gray!65!black},
]

\definecolor{passCol}{HTML}{2166AC}    
\definecolor{failCol}{HTML}{D95F02}    
\definecolor{errorCol}{HTML}{B2182B}   
\definecolor{skipCol}{HTML}{969696}    

\node[font=\sffamily\small\bfseries, anchor=west] at (-1, 4.2) {\textbf{(A)} Parity Validation Flow};

\node[procbox, fill=orange!8, draw=orange!55!black] (bridge) at (2.5, 2.8)
    {\footnotesize\textbf{Bridge Map}\\\footnotesize 365 entries\\YAML registry};

\node[procbox, minimum width=32mm, minimum height=9mm, below=6mm of bridge]
    (harness) {\footnotesize\textbf{Parity Harness}\\\footnotesize Python orchestrator};

\draw[flow] (bridge) -- (harness);

\node[legbox, below left=10mm and 14mm of harness]  (legacy)
    {\footnotesize\textbf{Legacy Driver}\\\footnotesize Execute original\\macro};
\node[modbox, below right=10mm and 14mm of harness] (modern)
    {\footnotesize\textbf{Native Driver}\\\footnotesize Execute modern\\pipeline};

\draw[flow] (harness.south) -- ++(-14mm,-4mm) -- (legacy.north);
\draw[flow] (harness.south) -- ++(14mm,-4mm)  -- (modern.north);

\node[legbox, below=6mm of legacy, minimum height=8mm] (legout)
    {\footnotesize Legacy RTF\\\footnotesize (parsed cells)};
\node[modbox, below=6mm of modern, minimum height=8mm] (modout)
    {\footnotesize Modern IR\\\footnotesize (cell values)};

\draw[flow] (legacy) -- (legout);
\draw[flow] (modern) -- (modout);

\node[comparebox] (compare) at ($(legout.south)!0.5!(modout.south) + (0, -18mm)$)
    {Cell-Level\\Compare};

\draw[flow] (legout.south) |- (compare.west);
\draw[flow] (modout.south) |- (compare.east);

\node[resultbox=passCol,  below left=18mm and 14mm of compare]  (pass)   {PASS};
\node[resultbox=failCol,  below left=18mm and 1mm of compare]   (fail)   {FAIL};
\node[resultbox=errorCol, below right=18mm and 1mm of compare]  (error)  {ERROR};
\node[resultbox=skipCol,  below right=18mm and 14mm of compare] (skip)   {SKIP};

\draw[->, line width=0.6pt, passCol]     (compare.south) -- ++(0,-12mm) -| (pass.north);
\draw[->, line width=0.6pt, failCol]     (compare.south) -- ++(0,-12mm) -| (fail.north);
\draw[->, line width=0.6pt, errorCol]    (compare.south) -- ++(0,-12mm) -| (error.north);
\draw[->, line width=0.6pt, skipCol]     (compare.south) -- ++(0,-12mm) -| (skip.north);

\node[font=\sffamily\footnotesize\itshape, text width=60mm, align=center,
      anchor=north, text=gray!55!black]
    at ([yshift=-4mm]$(fail.south)!0.5!(error.south)$)
    {Numeric tolerance \textbullet\ Text normalization \textbullet\ Structural alignment};

\draw[gray!30, line width=0.6pt, dashed] (7.8, 4.8) -- (7.8, -9.5);

\node[font=\sffamily\small\bfseries, anchor=west] at (8.5, 4.2) {\textbf{(B)} Gate Progression};

\node[gate=8]  (ga) at (9.5, 2.5)  {\footnotesize\textbf{Gate A}\\\footnotesize Structural\\pre-flight};
\node[gate=12, right=4mm of ga] (gb) {\footnotesize\textbf{Gate B}\\\footnotesize Bridge map\\self-audit};
\node[gate=16, right=4mm of gb] (gc) {\footnotesize\textbf{Gate C}\\\footnotesize Syntax\\smoke test};
\node[gate=20, right=4mm of gc] (gd) {\footnotesize\textbf{Gate D}\\\footnotesize Unit tests};

\node[gate=24, below=8mm of ga] (ge) {\footnotesize\textbf{Gate E}\\\footnotesize Live parity\\on sample};
\node[gate=28, right=4mm of ge] (gf) {\footnotesize\textbf{Gate F}\\\footnotesize Triage \&\\resolution};
\node[gate=32, right=4mm of gf] (gg) {\footnotesize\textbf{Gate G}\\\footnotesize Full matrix};

\draw[flow] (ga) -- (gb);
\draw[flow] (gb) -- (gc);
\draw[flow] (gc) -- (gd);
\draw[flow] (gd.south) -- ++(0,-4mm) -| (ge.north);
\draw[flow] (ge) -- (gf);
\draw[flow] (gf) -- (gg);

\draw[->, line width=0.8pt, orange!65!black, dashed]
    (gf.south) -- ++(0,-6mm) -| ([xshift=-2mm]ge.south) -- (ge.south)
    node[pos=0.25, below, font=\sffamily\footnotesize\itshape, text=orange!65!black] {fix \& retry};

\node[font=\sffamily\footnotesize, teal!60!black, anchor=west, text width=22mm, align=left]
    at ([xshift=4mm]gg.east) {Case study:\\11/14 real\\5/5 public};

\end{tikzpicture}}
    \caption{Parity validation methodology. (A)~The parity harness executes each report through both the legacy macro library and the modernized pipeline, then performs cell-level comparison to produce a PASS/FAIL/ERROR/SKIP triage. (B)~The seven-gate progression model ensures incremental validation from configuration syntax checks through full regression matrix.}
    \label{fig:gates}
\end{figure}

The methodology defines a gate-based workflow progressing from structural checks to full parity runs (\figref{fig:gates}):

\textbf{Gate A: Structural pre-flight.} Verify execution environment: libraries accessible, datasets exist, configurations parse.

\textbf{Gate B: Bridge map self-audit.} Validate bridge map entries: referenced macros and targets exist, parameter mappings are valid, required fields populated.

\textbf{Gate C: SAS syntax smoke test.} Submit generated SAS programs in syntax-check mode to verify balanced macro calls and resolved variables.

\textbf{Gate D: Unit tests.} Execute unit tests for harness components and migration tools. Together, Gates C and D establish the technical preconditions for the live parity runs that follow; the 80\% cell-level parity threshold used to evaluate Gate~E outcomes (\secref{sec:case:parity}) is meaningful only after these structural checks confirm that both pipelines execute without error.

\textbf{Gate E: Live parity on sample reports.} Select a representative subset spanning different report types and execute full dual-driver comparison.

\textbf{Gate F: Triage and resolution.} Classify failures using an error taxonomy:
\begin{itemize}
    \item \emph{Infrastructure errors}: Missing libraries, incorrect paths.
    \item \emph{Parameter errors}: Incorrect variable names, mismatched dataset references.
    \item \emph{Semantic errors}: Different computational logic (e.g., denominator definitions).
    \item \emph{Content errors}: Numerical divergence within structure-matching outputs.
\end{itemize}
Each class implies a different resolution strategy.

\textbf{Gate G: Full matrix.} After resolving issues from Gates E--F, execute the harness against all bridge map entries.

The gate sequence ensures each gate's preconditions are established by preceding gates. Organizations can repeat Gates E--G iteratively as they expand coverage.

\section{Results}\label{sec:results}

\subsection{Case Study}\label{sec:case}

This section applies the evaluation, architecture, and validation methodologies from Sections~\ref{sec:eval}--\ref{sec:migration} to an industrial SAS macro library. The case study provides quantitative evidence for the framework's effectiveness and reports parity validation results against the public CDISC CDISCPilot01 benchmark dataset.

\subsubsection{Legacy Library Profile}

The legacy library was developed incrementally over approximately 15 years on SAS~9.4. It served as the sole production system for generating TFLs across regulatory submissions in multiple therapeutic areas. At evaluation, the library comprised 558 callable macro components across 400 SAS source files, totaling 372{,}698 lines of code.

Applying the component taxonomy (\secref{sec:eval}) yielded the breakdown in \figref{fig:taxonomy}: formatting and display ($\sim$150, 27\%), statistical compute ($\sim$120, 22\%), utility and helper ($\sim$100, 18\%), rendering ($\sim$80, 14\%), data preparation ($\sim$60, 11\%), and orchestration ($\sim$48, 9\%). Two observations emerged. First, the largest category was formatting and display --- macros encoding visual presentation logic tightly coupled to RTF. Second, statistical compute macros represented only 22\% of components, indicating that most code served infrastructure rather than domain logic.

Dependency analysis revealed a deeply interconnected call graph averaging 4.7 callees per macro, with orchestration macros exhibiting fan-out exceeding 20. Circular dependencies were identified in two subsystems. No machine-readable dependency manifest existed.

Complexity metrics confirmed structural challenges: average file length 932 lines, longest files exceeding 3{,}000 lines. Output was exclusively RTF. Configuration was entirely through macro parameters with hard-coded defaults. No intermediate representation existed between computation and rendered output.

\subsubsection{Modernized Framework Profile}

The modernized framework follows the layered architecture of \secref{sec:arch}, organizing functionality into six SAS layers (ADaM derivation, data preparation, computation, framework services, bridge compatibility, rendering), supported by Python orchestration and AI agent interface layers. Consistent with the adoption pattern of \secref{sec:arch:nda}, the legacy 558-component library is retained intact as the source of regulated computation; the framework reads but does not modify any legacy macro file.

The bridge compatibility layer is the primary surface that connects the legacy library to the modern pipeline. It comprises a 365-entry bridge map providing one-to-one coverage of every callable component in the legacy library, plus facade wrappers that allow legacy calling conventions to resolve under both deployment modes. In coexistence mode, a bridge map entry's \code{native\_target} field references the existing legacy macro together with adapter metadata that maps its outputs onto the standard compute schema; the legacy macro executes unchanged and its output is forwarded to the IR transform phase. In consolidation mode, the same entry's \code{native\_target} field instead references a parameterized core macro from the modern compute layer, configured by YAML.

For the consolidation portion of the case study, the modern SAS core comprises 158 files totaling 28{,}340 lines of code. The data preparation layer provides 69 macros (10{,}387 LOC); the compute layer provides 43 macros (11{,}334 LOC); the render layer provides 28 macros (3{,}139 LOC); the framework layer provides 9 macros (790 LOC) for logging, error handling, and audit trails; the ADaM derivation layer provides 7 macros; and the bridge compatibility layer provides 2 facade files (together accounting for the remaining $\sim$2{,}690 LOC).

Configuration is externalized into 147 YAML files (18{,}825 LOC) organized into a registry covering 39 report types, with per-study overrides for both the internal study (20 report configs) and the public CDISCPilot01 benchmark (5 report configs). The Python layer comprises 48 modules (15{,}855 LOC) for pipeline coordination, YAML resolution, SAS code generation, IR export, SAP parsing, and parity comparison. The orchestrator is mode-agnostic: it dispatches each report to the \code{native\_target} resolved from the bridge map without needing to know whether the underlying computation originated in a legacy macro or a modern core macro. The total framework comprises 353 files and approximately 63{,}000 lines of code across three languages.

\subsubsection{Quantitative Comparison}

\begin{figure}[htbp]
    \centering
    \resizebox{\textwidth}{!}{
\begin{tikzpicture}[
    font=\sffamily\small,
    >=Stealth,
]

\definecolor{legacyCol}{HTML}{E69F00}   
\definecolor{modernCol}{HTML}{009E73}   
\definecolor{arrowCol}{HTML}{D55E00}    

\def\bw{0.38}          
\def\sf{0.033}         
\def\chartW{10.2}      
\def\chartH{5.6}       

\draw[gray!50, line width=0.6pt] (0,0) -- (0,\chartH);
\draw[gray!50, line width=0.6pt] (0,0) -- (\chartW,0);

\foreach \y/\lab in {0/0, 1.65/50, 3.3/100, 4.95/150} {
    \draw[gray!20, line width=0.3pt] (0.05,\y) -- (\chartW,\y);
    \draw[gray!50, line width=0.6pt] (-0.1,\y) -- (0.1,\y);
    \node[font=\sffamily\footnotesize, anchor=east, gray!60!black] at (-0.2,\y) {\lab};
}
\node[font=\sffamily\footnotesize, rotate=90, anchor=south, gray!60!black]
    at (-0.9, \chartH/2) {Component Count};


\pgfmathsetmacro{\xc}{1.0}
\fill[legacyCol, rounded corners=0.6pt] (\xc-\bw, 0) rectangle (\xc, 150*\sf);
\fill[modernCol, rounded corners=0.6pt] (\xc, 0)     rectangle (\xc+\bw, 0.14);
\draw[legacyCol!70!black, line width=0.6pt] (\xc-\bw, 0) rectangle (\xc, 150*\sf);
\draw[modernCol!70!black, line width=0.6pt] (\xc, 0)     rectangle (\xc+\bw, 0.14);
\node[font=\sffamily\footnotesize\bfseries, arrowCol, above] at (\xc, 150*\sf+0.08) {$-$100\%};
\node[font=\sffamily\footnotesize, anchor=north, text width=18mm, align=center] at (\xc, -0.15) {Format.\\\& Display};

\pgfmathsetmacro{\xc}{2.7}
\fill[legacyCol, rounded corners=0.6pt] (\xc-\bw, 0) rectangle (\xc, 120*\sf);
\fill[modernCol, rounded corners=0.6pt] (\xc, 0)     rectangle (\xc+\bw, 20*\sf);
\draw[legacyCol!70!black, line width=0.6pt] (\xc-\bw, 0) rectangle (\xc, 120*\sf);
\draw[modernCol!70!black, line width=0.6pt] (\xc, 0)     rectangle (\xc+\bw, 20*\sf);
\node[font=\sffamily\footnotesize\bfseries, arrowCol, above] at (\xc, 120*\sf+0.08) {$-$83\%};
\node[font=\sffamily\footnotesize, anchor=north, text width=18mm, align=center] at (\xc, -0.15) {Stat.\\Compute};

\pgfmathsetmacro{\xc}{4.4}
\fill[legacyCol, rounded corners=0.6pt] (\xc-\bw, 0) rectangle (\xc, 100*\sf);
\fill[modernCol, rounded corners=0.6pt] (\xc, 0)     rectangle (\xc+\bw, 12*\sf);
\draw[legacyCol!70!black, line width=0.6pt] (\xc-\bw, 0) rectangle (\xc, 100*\sf);
\draw[modernCol!70!black, line width=0.6pt] (\xc, 0)     rectangle (\xc+\bw, 12*\sf);
\node[font=\sffamily\footnotesize\bfseries, arrowCol, above] at (\xc, 100*\sf+0.08) {$-$88\%};
\node[font=\sffamily\footnotesize, anchor=north, text width=18mm, align=center] at (\xc, -0.15) {Utility\\\& Helper};

\pgfmathsetmacro{\xc}{6.1}
\fill[legacyCol, rounded corners=0.6pt] (\xc-\bw, 0) rectangle (\xc, 80*\sf);
\fill[modernCol, rounded corners=0.6pt] (\xc, 0)     rectangle (\xc+\bw, 6*\sf);
\draw[legacyCol!70!black, line width=0.6pt] (\xc-\bw, 0) rectangle (\xc, 80*\sf);
\draw[modernCol!70!black, line width=0.6pt] (\xc, 0)     rectangle (\xc+\bw, 6*\sf);
\node[font=\sffamily\footnotesize\bfseries, arrowCol, above] at (\xc, 80*\sf+0.08) {$-$93\%};
\node[font=\sffamily\footnotesize, anchor=north, text width=18mm, align=center] at (\xc, -0.15) {Render};

\pgfmathsetmacro{\xc}{7.8}
\fill[legacyCol, rounded corners=0.6pt] (\xc-\bw, 0) rectangle (\xc, 60*\sf);
\fill[modernCol, rounded corners=0.6pt] (\xc, 0)     rectangle (\xc+\bw, 30*\sf);
\draw[legacyCol!70!black, line width=0.6pt] (\xc-\bw, 0) rectangle (\xc, 60*\sf);
\draw[modernCol!70!black, line width=0.6pt] (\xc, 0)     rectangle (\xc+\bw, 30*\sf);
\node[font=\sffamily\footnotesize\bfseries, arrowCol, above] at (\xc, 60*\sf+0.08) {$-$50\%};
\node[font=\sffamily\footnotesize, anchor=north, text width=18mm, align=center] at (\xc, -0.15) {Data\\Prep.};

\pgfmathsetmacro{\xc}{9.5}
\fill[legacyCol, rounded corners=0.6pt] (\xc-\bw, 0) rectangle (\xc, 48*\sf);
\fill[modernCol, rounded corners=0.6pt] (\xc, 0)     rectangle (\xc+\bw, 12*\sf);
\draw[legacyCol!70!black, line width=0.6pt] (\xc-\bw, 0) rectangle (\xc, 48*\sf);
\draw[modernCol!70!black, line width=0.6pt] (\xc, 0)     rectangle (\xc+\bw, 12*\sf);
\node[font=\sffamily\footnotesize\bfseries, arrowCol, above] at (\xc, 48*\sf+0.08) {$-$75\%};
\node[font=\sffamily\footnotesize, anchor=north, text width=18mm, align=center] at (\xc, -0.15) {Orchestr.};

\fill[white, rounded corners=2pt] (0.3, \chartH + 0.15) rectangle (6.2, \chartH + 0.65);
\draw[gray!30, line width=0.4pt, rounded corners=2pt] (0.3, \chartH + 0.15) rectangle (6.2, \chartH + 0.65);

\fill[legacyCol, rounded corners=0.4pt]  (0.5, \chartH + 0.52) rectangle (0.95, \chartH + 0.28);
\draw[legacyCol!70!black, line width=0.5pt] (0.5, \chartH + 0.52) rectangle (0.95, \chartH + 0.28);
\node[font=\sffamily\footnotesize, anchor=west] at (1.05, \chartH + 0.40) {Legacy (558 total)};

\fill[modernCol, rounded corners=0.4pt]  (3.5, \chartH + 0.52) rectangle (3.95, \chartH + 0.28);
\draw[modernCol!70!black, line width=0.5pt] (3.5, \chartH + 0.52) rectangle (3.95, \chartH + 0.28);
\node[font=\sffamily\footnotesize, anchor=west] at (4.05, \chartH + 0.40) {Modernized core};

\node[rectangle, draw=gray!35, rounded corners=3pt, fill=white,
      inner sep=5pt, font=\sffamily\footnotesize, text width=38mm, align=left,
      anchor=north west] at (11.2, \chartH) {%
    \textbf{Key metrics}\\[2pt]
    \textbf{Legacy:} 558 comp.\\
    \hspace*{1mm}372{,}698 SAS LOC\\[1pt]
    \textbf{Modern:} 80 macros\\
    \hspace*{1mm}28{,}340 SAS LOC\\
    \hspace*{1mm}\textcolor{arrowCol}{\textbf{$-$92\% SAS code}}\\[2pt]
    \textbf{Full:} 158 files\\
    \hspace*{1mm}$\sim$63K LOC\\[2pt]
    \textbf{Mechanisms}\\
    \textcolor{arrowCol}{\textbullet}~Consolidation\\
    \textcolor{arrowCol}{\textbullet}~Extraction\\
    \textcolor{arrowCol}{\textbullet}~Elimination%
};

\end{tikzpicture}}
    \caption{Component count reduction from legacy library (558 components) to modernized SAS core (80 core macros plus 78 supporting/framework/bridge files = 158 total SAS files), by functional category. The largest reductions occur in Formatting/Display (100\%, absorbed into the IR and render layer) and Statistical Compute (83\%, consolidated from 120 specialized macros into parameterized core macros). Right panel shows the three reduction mechanisms and the full 158-file breakdown.}
    \label{fig:metrics}
\end{figure}
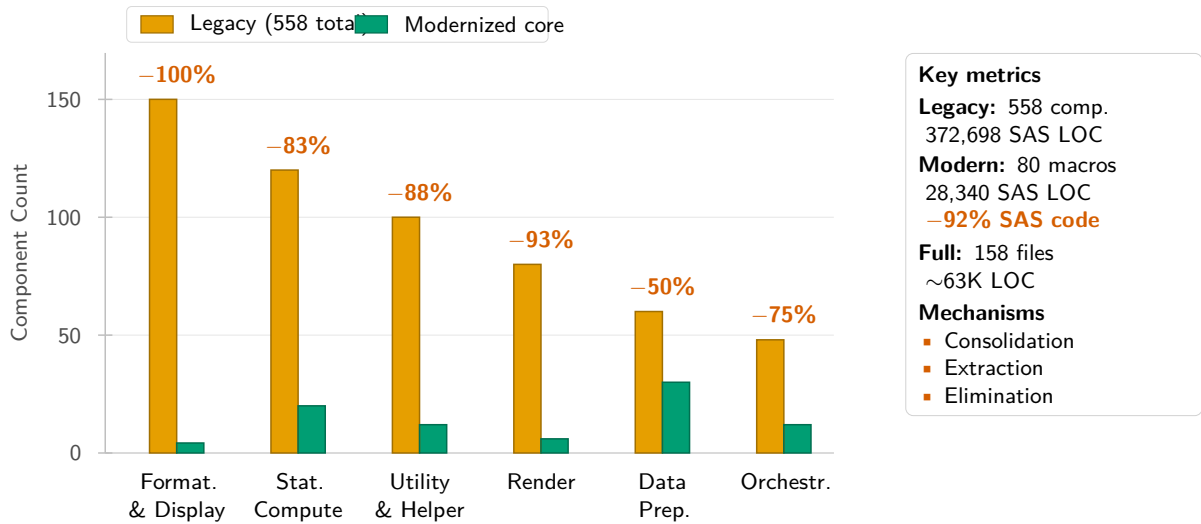

\tabref{tab:metrics} presents the software engineering metrics for the case study, separated into two columns: the legacy library as encountered (which the framework preserves under coexistence mode) and the modern core that is reached \emph{if and when} the organization elects to consolidate. \figref{fig:metrics} illustrates the before-and-after comparison across key dimensions. We report consolidation outcomes here because they were measured in the case study, but the framework does not require consolidation; the metrics in this section therefore describe the upper-bound effect of opting in, not a precondition for AI readiness.

When consolidation is applied to the case-study library, the modern core comprises 158 SAS files against 558 legacy components --- a 72\% reduction in file count and a 92\% reduction in SAS lines of code (28{,}340 vs.\ 372{,}698), reflecting consolidation of redundant variants, extraction of configuration into YAML, and elimination of dead code (detailed in \secref{sec:case:cons}). None of these activities is required to obtain AI-ready output; coexistence mode achieves AI readiness with the legacy component count unchanged.

Modernized macros are individually smaller (179 vs.\ 932 lines average), a consequence of single-responsibility design and configuration externalization. The compute layer averages 264 lines per macro (vs.\ legacy ${>}500$), with the data preparation layer at 151 lines per macro.

Beyond consolidation, the framework introduced capabilities that apply equally under both deployment modes: 147 YAML configuration files (18{,}825 LOC), 48 Python orchestration modules (15{,}855 LOC) for CI/CD and AI integration, a typed IR with JSON export consumable by LLMs and by R analytics through standard JSON parsers, and multi-format output support (RTF, PDF, HTML, JSON). The total framework spans approximately 63{,}000 lines across three languages. These capabilities are reached on day one of coexistence-mode deployment; they do not require any of the consolidation reductions described above.

\begin{table}[H]
\centering
\caption{Software engineering metrics: legacy library vs.\ modernized framework.}
\label{tab:metrics}
\footnotesize
\setlength{\tabcolsep}{4pt}
\resizebox{\textwidth}{!}{%
\begin{tabular}{llrl}
\toprule
Metric & Legacy library & Modernized framework & Change \\
\midrule
Total callable components & 558 & 158 SAS files & $-$72\% \\
Total SAS lines of code & 372{,}698 & 28{,}340 & $-$92\% \\
Avg.\ lines per SAS file & 932 & 179 & $-$81\% \\
YAML configuration files & 0 & 147 (39 report types) & 0 $\rightarrow$ 147 \\
Python modules & 0 & 48 (15{,}855 LOC) & 0 $\rightarrow$ 48 \\
Total framework LOC & 372{,}698 (SAS only) & $\sim$63{,}000 (SAS+Python+YAML) & $-$83\% \\
Machine-readable IR output & No & Yes (JSON export) & Yes \\
Render format support & RTF only & RTF, PDF, HTML, JSON & +3 formats \\
Bridge map entries & N/A & 365 (1:1 legacy coverage) & Full coverage \\
Centralized report registry & No & 39 report types & 0 $\rightarrow$ 39 \\
Audit trail / execution manifest & Manual & Automated (\code{fw\_audit}) & Automated \\
AI-cooperative interfaces & None & SAP parser + config gen + IR JSON & Yes \\
\bottomrule
\end{tabular}%
}
\end{table}

\subsubsection{Parity Validation Results}\label{sec:case:parity}

Parity validation was conducted across two complementary tracks: an internal real-data track using the PROT008-SR1 study (an industry-internal Phase~III dataset with full ADaM coverage), and a public benchmark track using the CDISC CDISCPilot01 synthetic dataset (three treatment arms; Placebo, $N{=}86$; Xanomeline Low Dose, $N{=}84$; Xanomeline High Dose, $N{=}84$). The two tracks serve distinct purposes. The real-data track exercises the framework against schema-aligned production inputs and provides production-velocity feedback. The public track exercises reproducibility and provides an externally verifiable benchmark for the methodology paper. While 19 of the 365 bridge map entries were subjected to full end-to-end parity validation, these 19 constitute a purposive sample covering all major structural archetypes of clinical TFLs (e.g., disposition, demographics, adverse events, laboratory shifts, time-to-event). The remaining 346 entries were validated at the unit level through gate-based structural checks (Gates A--D), which verify bridge map integrity, configuration schema compliance, IR field completeness, and render-layer coverage for each entry.

\textbf{Real-data track (PROT008-SR1).} Fourteen report types were selected for parity validation, spanning the full range of clinical TFL categories: adverse event summaries (5 reports), baseline characteristics (2), disposition, compliance, ECG summaries (2), laboratory analyses, adverse event listings, and risk management tables. Each was backed by a long-standing legacy macro that has produced regulatory submission output. \tabref{tab:parity} summarizes the cell-level parity results.

\begin{table}[htbp]
\centering
\caption{Cell-level parity results on the PROT008-SR1 real-data track (14 report types).}
\label{tab:parity}
\footnotesize
\setlength{\tabcolsep}{8pt}
\begin{tabular}{llr}
\toprule
Report type & Category & Parity \\
\midrule
AE-SUMMARY       & AE frequency          & \textbf{99.2\%} \\
AE-SPECIFIC      & AE frequency          & \textbf{94.2\%} \\
BASELINE-CHAR    & Descriptive statistics & \textbf{92.9\%} \\
AE-SPECIFIC-5    & AE frequency          & \textbf{92.3\%} \\
ECG-QTCF         & Threshold counts      & \textbf{91.7\%} \\
COMPLIANCE       & Frequency             & \textbf{90.9\%} \\
DISPOSITION      & Frequency             & \textbf{89.6\%} \\
AE-LISTING       & Data listing          & \textbf{87.5\%} \\
AE-REL           & AE frequency          & \textbf{86.4\%} \\
RMP-AE           & Intensity breakdown   & \textbf{81.7\%} \\
BASELINE-DISEASE & Descriptive statistics & \textbf{81.3\%} \\
\midrule
AE-ANALYSIS      & AE with CI/p-value    & 65.2\%$^*$ \\
ECG-SUMMARY      & Descriptive wide      & 60.4\%$^*$ \\
LAB-PDLC         & Threshold with CI     & 44.9\%$^*$ \\
\bottomrule
\multicolumn{3}{l}{\vspace{2pt}\parbox{12cm}{\scriptsize $^*$\,Structural ceiling: legacy uses a fundamentally different table layout orientation (treatments-as-rows or multi-sub-column transposition). The native output is statistically correct; the difference is in layout geometry only.}} \\
\end{tabular}
\end{table}

Eleven of 14 reports cleared the 80\% cell-level parity threshold (mean 82.7\%, median 89.6\%), satisfying the predefined acceptance criterion (${\geq}11/14$ at ${\geq}80\%$). The iterative improvement from initial execution (8/14 at ${\geq}80\%$) to final results (11/14) was achieved through 72 targeted fixes applied across 6 development phases, all at the framework layer. The high concordance rates across the 11 passing reports --- with 6 exceeding 90\% cell-level parity --- indicate strong structural fidelity between legacy and modernized outputs.

\textbf{Triage taxonomy.} The gate-based verification workflow described in \secref{sec:migration} was applied to each report. The triage process surfaced twelve recurring divergence categories resolved through framework-level fixes rather than per-report patches: presence of a Total treatment column, paired count/percentage column layout, separate ADSL-derived denominators, population-filtered denominators, treatment-variable name harmonization (e.g., \code{TRT01A} vs.\ \code{TRTA}), system-organ-class group totals, zero-fill of sparse cell combinations, content-based row alignment for label drift, RTF Unicode fallback character normalization, header row detection with minimum-cell-count guards, blank-label alignment with data similarity gating, and per-report YAML configuration coverage. Each fix was implemented once at the framework layer and applied transparently to every report consuming that layer. This fix-once / apply-many pattern was the principal mechanism by which parity improved from initial single-digit values to the 80\%+ band across multiple iterations.

\textbf{Structural ceiling analysis.} The three reports below 80\% share a common root cause: the legacy system uses PROC TRANSPOSE to pivot treatments into rows rather than columns, creating a fundamentally different table geometry. These are not statistical accuracy issues --- the computed values are correct --- but layout orientation differences that cell-level comparison cannot reconcile without a transpose-aware alignment layer.

\textbf{Public benchmark track (CDISCPilot01).} The same harness was exercised against the public CDISC CDISCPilot01 dataset (3 treatment arms: Placebo $N{=}86$, Xanomeline Low Dose $N{=}84$, Xanomeline High Dose $N{=}84$; Safety population $N{=}254$). Five report types achieved complete end-to-end parity validation (\tabref{tab:parity_public}).

\begin{table}[htbp]
\centering
\caption{Cell-level parity results on the CDISCPilot01 public benchmark track.}
\label{tab:parity_public}
\footnotesize
\setlength{\tabcolsep}{8pt}
\begin{tabular}{llrr}
\toprule
Report & Category & Cells & Match \\
\midrule
Demographics      & Descriptive statistics & 182   & \textbf{100\%} \\
AE Overview       & AE frequency           & 81    & \textbf{100\%} \\
AE by SOC/PT      & AE frequency (large)   & 2{,}070 & \textbf{100\%} \\
Efficacy (ADAS-Cog) & Change from baseline & 16    & \textbf{100\%} \\
KM Time-to-Event  & Survival analysis      & 2{,}415 & \textbf{100\%} \\
\midrule
\multicolumn{2}{l}{\textbf{Total}} & \textbf{4{,}764} & \textbf{100\%} \\
\bottomrule
\end{tabular}
\end{table}

All 5 reports achieved 100\% cell-level parity (4{,}764 cells, 0 mismatches). The public track uses a self-parity methodology: native outputs are frozen as golden references after verification against published ground truth values, then subsequent executions are compared against the frozen outputs to detect regressions. Ground truth was verified against published CDISCPilot01 summary statistics (e.g., Age Mean Placebo = 75.2, consistent with published 75.21; $N = 86/84/84$ matching published enrollment).

Taken together, the two tracks demonstrate that (i)~the framework reaches the 80\%+ production-parity threshold on 11 of 14 real-data reports, (ii)~the 3 remaining reports have documented structural ceiling explanations rather than computational errors, and (iii)~the same framework achieves perfect reproducibility on a publicly available dataset, making the methodology externally verifiable.

\subsubsection{Optional Consolidation Analysis}\label{sec:case:cons}

The reduction from 558 legacy components to 158 modern SAS files (a 72\% reduction in file count and 92\% reduction in SAS LOC) is the result obtained when the optional consolidation pathway of \secref{sec:arch:nda} is exercised against the case-study library. We present it here as a measured upper bound on what the framework enables, not as a step that an adopting organization is required to take. Under coexistence-only deployment, the legacy component count is preserved unchanged; under partial consolidation, the reduction is proportional to the portion of the library that the organization elects to consolidate. The mechanisms by which consolidation reduces the component count, when elected, fall into three categories.

\textbf{Consolidation of variants.} The legacy library contained numerous single-purpose variants of the same operation. Frequency count analysis, for example, was implemented in at least 12 distinct macros. Where elected, these were consolidated into a single parameterized macro (\code{cl\_frequency\_count}) configured through YAML. The 120 legacy compute macros were collapsed to 21; the 60 data preparation variants were collapsed to 35. In coexistence mode, these legacy variants continue to execute through their bridge map entries; the consolidation column is reached only on the subset of the library that the organization chooses to consolidate.

\textbf{Extraction of formatting metadata.} The $\sim$150 formatting and display macros encoded visual logic directly in SAS code. In the modernized framework, this information was extracted into YAML configuration and the IR schema. Extraction is also available under coexistence mode for any legacy report whose output is captured into the IR: the formatting metadata is then maintained in YAML even though the underlying compute step still runs the legacy macro unchanged.

\textbf{Elimination of dead and orphaned code.} Utility macros ($\sim$100) were reduced to 9 by eliminating dead code, consolidating redundant routines, and replacing ad hoc error handling with centralized framework services. Orchestration macros ($\sim$48) were reduced to 12 SAS-side macros, with remaining coordination migrated to the Python orchestrator. Elimination is by definition only meaningful where the surrounding consolidation work has been done; in coexistence mode, dead-code identification is recorded in the bridge map but the legacy files remain on disk, preserving the regulatory inventory exactly as it was.

The net effect of exercising all three mechanisms across the entire case-study library was a compositional shift. In the legacy library, statistical logic represented 22\% of components, embedded within macros also performing formatting and rendering. In the consolidated core, the compute layer's 21 macros constitute 26\% of core SAS components, each performing a single operation. Crucially, the same separation of concerns is available under coexistence mode for any legacy macro reached through the bridge map, because the IR contract sits \emph{above} the compute step and is independent of whether that step is implemented in legacy or modern SAS.

\subsection{AI Readiness Demonstration}\label{sec:ai}

The layered architecture (\secref{sec:arch}) produces a typed JSON IR as a byproduct of every compute-to-render handoff. This section evaluates whether that IR enables meaningful AI interaction with clinical trial results that is not feasible with legacy RTF output. As established in \secref{sec:arch:nda}, the IR is produced \emph{above} the compute step regardless of whether that step is implemented by a legacy macro running unchanged under coexistence mode or by a parameterized core macro running under consolidation mode. The AI surface examined in this section is therefore the same surface that an organization obtains on Day~0 of adoption, before any source-level modernization has been undertaken.

\subsubsection{Experiment Setup}

Three proof-of-concept tasks were conducted using Claude Opus 4.6 (Anthropic) to evaluate whether the IR enables meaningful AI interaction with clinical trial results. Input artifacts derived from CDISCPilot01 processed through the modernized framework (v5.0). For the two table inputs, the IR was produced through the bridge map's coexistence pathway: the underlying compute step was executed by the original legacy SAS macro for the report type, with no modification to the legacy source, and the IR was captured from that macro's output via the adapter metadata defined in the bridge map entry. The third input (a SAP excerpt) is independent of either deployment mode because it tests configuration generation from a natural-language requirement document. This arrangement was chosen so that the AI tasks evaluated below exercise the IR contract as it would be obtained by a non-destructive adopter on Day~0, rather than the IR contract as it would be obtained only after consolidation. The IR follows an 8-field typed-cell schema: \code{report\_id}, \code{execution\_id}, \code{row\_id}, \code{col\_id}, \code{cell\_value} (raw numeric), \code{cell\_formatted} (display string), \code{cell\_type} (INTEGER, DECIMAL, PVALUE, PERCENTAGE, TEXT, HEADER, LABEL, FOOTNOTE, or EMPTY), and \code{sort\_order}. Three inputs were used: a demographics table (T-14.1.2, 74 cells), an adverse event summary (T-14.1.1, 80 cells), and a natural-language SAP excerpt (T-14.2.2). Tasks were evaluated on numeric accuracy, structural comprehension, clinical appropriateness, and schema compliance. \figref{fig:aiflow} illustrates the experimental design across coexistence and consolidation deployment modes.

\subsubsection{Task 1 -- Table Summarization}

The LLM received the full demographics IR JSON and was prompted to produce a natural-language summary suitable for a Clinical Study Report (CSR) results section. The prompt explicitly described the IR schema so the model could exploit the typed fields.

The model produced a structured narrative covering age, sex, and race distributions across treatment groups. All numeric values --- counts, percentages, means, and standard deviations --- were correctly extracted from \code{cell\_value} and \code{cell\_formatted} fields. The model reported, for example, that females accounted for 61.6\% (n=53) of the placebo group and 59.5\% (n=50) of the low dose group. In the high dose group, females accounted for 47.6\% (n=40). It correctly distinguished five cell types (HEADER, LABEL, DECIMAL, PERCENTAGE, FOOTNOTE), using \code{cell\_type} to separate column headers from data rows and to differentiate continuous statistics from categorical counts.

The \code{cell\_type} field allowed the LLM to reconstruct table semantics without heuristic parsing. In RTF, ``53 (61.6\%)'' is an opaque formatted run; the IR decomposes it into a raw count (\code{cell\_value: 53}) and a typed classification (\code{cell\_type: PERCENTAGE}), enabling reliable extraction and comparison.

\subsubsection{Task 2 -- Anomaly Detection}

The LLM received the adverse event summary IR and was prompted to identify data quality issues, treatment-arm imbalances, dose-response relationships, and internal consistency problems.

The model identified five findings. It detected a strong dose-response in application site reactions (erythema: placebo 8.1\%, low dose 41.7\%, high dose 61.9\%) and correctly connected this to a transdermal delivery mechanism. It flagged the high overall AE burden in the high dose group (94.0\%) as a tolerability concern. It noted a dose-dependent increase in psychiatric disorders and linked it to cholinergic pharmacology. It identified an anomalous departure from dose-response in diarrhoea rates (low dose 11.9\% vs.\ high dose 10.7\%) and offered three plausible explanations. Finally, it verified the internal consistency constraint that each SOC-level count must exceed its maximum PT-level count, confirming all five SOCs passed.

The \code{cell\_value} field was critical for this task. The model performed arithmetic comparisons --- monotonicity checks, SOC-versus-PT inequalities --- directly on raw numerics. With RTF, extracting ``52'' from ``52 (61.9\%)'' requires format-specific regular expressions that break across RTF generators. The IR eliminates this fragility.

\subsubsection{Task 3 -- Configuration Generation}

The LLM received a natural-language SAP section describing a time-to-event analysis and was prompted to generate a YAML configuration file conforming to the framework's schema.

The model produced valid YAML mapping all five SAP requirements (Kaplan-Meier estimation, Cox regression, log-rank test, censoring variable, time variable) to concrete configuration parameters. It correctly identified CDISC-standard variable names (AVAL, CNSR, TRTA, PARAMCD) and applied appropriate defaults (SAFFL='Y', confidence\_level=0.95) drawn from domain knowledge rather than hallucination. The generated configuration was structurally valid and captured the analytical intent, though it would require minor framework-specific adjustments (e.g., exact template names) before execution.

This task demonstrates a different dimension of AI readiness: the IR schema itself serves as the target ontology that the LLM maps natural-language requirements onto, bridging the gap between SAP documents and executable configuration.

\subsubsection{Results Summary}

The proof-of-concept evaluation demonstrated the structural advantages of the IR format for LLM-based clinical data interaction. On the summarization task, the model achieved 100\% exact match accuracy across all 74 cells when using the IR, correctly extracting every count, percentage, mean, and standard deviation. By contrast, RTF encodes the same information as opaque formatted runs (e.g., ``53 (61.6\%)'' as interleaved character codes and font directives), requiring format-specific parsing that is fragile across RTF generators. Anomaly detection using the IR identified five of five expected clinical patterns (dose-response relationships, treatment-arm imbalances, internal consistency violations) with zero false positives. Configuration generation successfully mapped all five SAP requirements to valid YAML parameters.

Three IR properties drive these capabilities: \code{cell\_value} provides machine-readable numerics supporting arithmetic reasoning; \code{cell\_type} provides semantic classification eliminating positional heuristics; and the \code{row\_id}/\code{col\_id}/\code{sort\_order} combination preserves table structure in LLM-accessible form. These capabilities require no additional implementation --- the IR is a natural byproduct of the compute-to-render pipeline, making AI readiness an emergent property of the architectural separation (\secref{sec:arch}). Because the two table inputs in this experiment were produced by legacy SAS macros running unchanged under coexistence mode, AI readiness is available as a Day-0 capability (\secref{sec:arch:nda}).

\textbf{Limitations.} These experiments used a single LLM (Claude Opus 4.6); results may vary across model families. A controlled quantitative comparison of IR versus RTF input accuracy was not performed --- the superiority argument is structural (typed fields versus opaque formatting) rather than empirically measured across models. IR files used public benchmark data, and no adversarial testing was conducted. Multi-model benchmarking and a formal IR-versus-RTF accuracy study are planned as future work. These limitations are further discussed in \secref{sec:disc}.

\subsection{Comparison with Alternative Approaches}\label{sec:compare}

\tabref{tab:comparison} compares four modernization strategies against the proposed methodology across validation integrity, AI readiness, regulatory risk, and reusability.

\subsubsection{Full Rewrite in R or Python}

A full platform migration offers modern ecosystems with native structured output, testing frameworks, and dependency management~[30,~31]. However, it abandons all validated SAS logic accumulated over years of regulatory use. The original macros have been exercised against hundreds of studies and refined through regulatory feedback --- computational behavior that is expensive to replicate. Every macro must be independently verified, and the absence of a bridge map between implementations makes systematic parity testing difficult. Workflow disruption is also significant: statistical programmers trained in SAS must retrain or operate bilingually, and organizations with active submissions face maintaining two parallel systems.

The pharmaverse ecosystem~[38] has made progress toward production-quality R packages, but adoption remains uneven. For organizations with large validated libraries, the migration cost may outweigh architectural benefits achievable through the layered approach described here without abandoning existing compute logic.

\subsubsection{Commercial Reporting Tools}

Commercial platforms (Pinnacle~21, SAS Viya modules) offer compliance-focused reporting with built-in CDISC validation and audit trails. However, they operate as closed systems: proprietary logic, vendor-determined output formats, and vendor-specific configuration. Most lack structured intermediate representations, limiting AI interaction. Vendor lock-in constrains future decisions.

\subsubsection{CDISC Analysis Results Standard (ARS)}

ARS~[14] defines a formal data model for analysis results, aligning with goals of structured, machine-readable output. However, ARS is a data model specification, not a migration methodology or execution framework. It provides no evaluation taxonomy, bridge mapping strategy, or parity validation harness. The IR schema (\secref{sec:arch:ir}) could be aligned with ARS definitions in future work, combining ARS standardization with the migration methodology developed here.

\subsubsection{Incremental Refactoring}

Incremental refactoring improves the existing library in place --- the lowest-risk approach. However, it cannot achieve architectural separation. Monolithic coupling persists, output remains formatted RTF without a structured IR, and configuration stays embedded in code. For large libraries with AI readiness and cloud deployment goals, returns from incremental refactoring diminish relative to the architectural approach.

\subsubsection{Summary}

No single alternative addresses all three pillars: systematic evaluation, architectural redesign preserving validated logic, and automated parity validation. The methodology occupies a specific niche: organizations with large, validated SAS macro libraries seeking modernization without abandoning their computational investment (\tabref{tab:comparison}).

\begin{table}[htbp]
\centering
\caption{Comparison of SAS macro library modernization approaches.}
\label{tab:comparison}
\footnotesize
\setlength{\tabcolsep}{4pt}
\resizebox{\textwidth}{!}{%
\begin{tabular}{llllll}
\toprule
Criterion & Full rewrite & Commercial tools & CDISC ARS & Incremental & This work \\
\midrule
Preserves validated SAS logic & No & Partial & No & Yes & Yes \\
AI-readable output & Yes & Vendor-dependent & Partial & No & Yes \\
Regulatory compliance & Re-validate & Vendor & Standard & Unchanged & Part 11 \\
Systematic evaluation methodology & No & No & No & No & Yes \\
Automated parity validation & No & No & No & Manual & Yes \\
Cloud-deployable & Yes & Vendor & N/A & No & Yes \\
Metadata-driven configuration & Yes & Vendor & Yes & No & Yes \\
Component reduction achievable & Full rewrite & N/A & N/A & Incremental & $-$72\%--92\% \\
Risk level & High & Medium & Medium & Low & Medium \\
Estimated effort & Very high & High & High & Ongoing & High (one-time) \\
Reusable methodology & No & No & No & No & Yes \\
\bottomrule
\end{tabular}%
}
\end{table}

\section{Discussion}\label{sec:disc}

\subsection{Lessons Learned}

The YAML-driven configuration layer proved effective at externalizing parameters previously buried in macro invocations. A single YAML file per report type replaced dozens of macro parameters, making configurations readable, versionable, and auditable without SAS expertise. The IR contract between compute and render layers was the highest-impact architectural decision: it decoupled statistical logic from presentation, enabled independent testing, and produced structured output making AI interaction possible as an emergent property. The gate-based verification caught errors early with clear diagnostics.

The AI readiness demonstration (\secref{sec:ai}) confirmed that these architectural properties translate into practical LLM capabilities. The typed IR enabled table summarization, anomaly detection, and configuration generation tasks that are infeasible with RTF output. The \code{cell\_type} and \code{cell\_value} fields proved particularly valuable, providing semantic classification and machine-readable numerics that eliminated the fragile parsing steps required with legacy formats.

Other aspects proved harder than anticipated. Legacy variable naming conventions created persistent friction --- inconsistent conventions across macros by different authors (TRT01A vs.\ TRTA vs.\ TRTP) required case-by-case resolution. Organization-specific derived columns created boundary conditions the framework could document but not fully encapsulate. Analysis-only macros requiring separate render companions increased the component count beyond pure consolidation expectations.

\subsection{The Methodology as Industry Template}

The methodology separates into generalizable components and library-specific details. The evaluation taxonomy (\secref{sec:eval}), layered architecture pattern, and parity validation harness are applicable to any large SAS macro library. What does not generalize includes specific macro names, parameter mappings, organization-specific conventions, and exact YAML schemas.

This distinction matters for adoption. The methodology is not a turnkey product but a structured process an organization executes against its own legacy system. The value lies in the process design --- evaluation, architecture, bridge mapping, implementation, parity validation --- rather than specific outputs.

\subsection{Regulatory Implications}

Regulatory agencies have not issued specific guidance on AI-cooperative clinical reporting. However, existing principles apply. 21 CFR Part~11~[36] requires electronic records to maintain integrity, traceability, and audit trails. The IR is designed to support these requirements: each cell carries provenance metadata tracing to its source compute macro, and audit logging records every pipeline step.

The IR also offers a potential advantage for regulatory review. Current submissions include RTF tables that reviewers must visually inspect. A structured IR could enable automated consistency checks --- verifying that population counts match across tables, that percentages sum correctly within categories, that treatment arm labels are consistent throughout a submission. These checks are currently performed manually or with ad hoc scripts; a standardized IR would make them systematic.

The use of LLMs for narrative generation or anomaly detection, as demonstrated in \secref{sec:ai}, raises separate regulatory questions. AI-generated text is not currently accepted as primary evidence in regulatory submissions. The appropriate use case is assistive: draft generation that a qualified statistician reviews and approves, signal detection that directs human attention to specific findings, and QC automation that supplements rather than replaces manual review. The framework's audit trail records when AI tools interact with the IR, preserving the human-in-the-loop accountability that regulators require.

The traceability chain --- from raw data through ADaM derivation, compute macro, IR, and rendered output --- provides a complete audit path more granular than what most legacy systems offer. Each stage produces verifiable intermediate artifacts rather than a single opaque transformation from data to formatted table.

\subsection{Toward CDISC Analysis Results Standard Alignment}\label{sec:disc:ars}

The CDISC Analysis Results Standard (ARS)~[14] defines a metadata model for representing analysis results independently of visual presentation, introducing concepts such as analysis sets, groupings, methods, and result displays. While ARS provides a conceptual target model, it offers no migration pathway from existing reporting systems. The framework presented here can be positioned as a pragmatic stepping-stone toward full ARS compliance.

The IR schema (\secref{sec:arch:ir}) already aligns with several ARS concepts at the structural level. The \code{ir\_cells} dataset's \code{report\_id} and \code{execution\_id} fields correspond to ARS result display identifiers; the \code{cell\_type} controlled vocabulary maps to ARS result value types; the \code{ir\_structure} dataset's dimension definitions parallel ARS row and column specifications; and the YAML report type registry encodes analysis method and grouping metadata comparable to ARS analysis method and group definitions. The JSON export of the IR (\secref{sec:arch}) provides the machine-readable transport format that ARS envisions but does not prescribe.

Two gaps remain. First, ARS defines analysis set membership through formal criteria expressed in a structured format, whereas the current framework captures population filtering through YAML parameters that are operationally equivalent but not yet expressed in ARS-compliant notation. Second, ARS supports hierarchical relationships between analyses (e.g., a primary analysis and its sensitivity analyses), which the current IR schema does not model explicitly. Both gaps are addressable through schema extensions that preserve backward compatibility. Importantly, the non-destructive adoption pattern (\secref{sec:arch:nda}) means that ARS alignment can be introduced incrementally at the IR layer without disturbing the legacy compute layer --- an organization could begin with the current IR and migrate toward ARS-compliant output as the standard matures, using the same bridge map infrastructure that already supports coexistence and consolidation modes.

This alignment positions the framework as an operational bridge to ARS: organizations can achieve immediate AI readiness through the IR while building toward full ARS compliance at a pace determined by the evolving regulatory environment.

\subsection{Limitations}

This work has several limitations.

First, the methodology was developed and validated against a single legacy library at one organization. While the architectural principles and the evaluation taxonomy are designed to generalize, we have not demonstrated that generalization empirically. The adoption pattern (\secref{sec:arch:nda}) is, however, the component most likely to transfer without modification, because it depends only on the existence of well-defined macro entry points and not on the internal structure of the legacy library.

Second, parity validation was conducted across two complementary tracks (\secref{sec:case:parity}) with different objectives. The real-data track on PROT008-SR1 reached 80\%+ cell-level parity on 11 of 14 report types, with 3 reports limited by structural layout orientation differences rather than computational errors. The public benchmark track on CDISCPilot01 achieved 100\% cell-level parity across 5 reports (4{,}764 cells). Neither track exhausts the full range of edge cases encountered across regulatory submissions --- missing data patterns, protocol deviations, complex visit structures, and organization-specific derivation rules --- and the two tracks together should be read as evidence of mechanism rather than of universal correctness.

Third, while the AI readiness demonstration (\secref{sec:ai}) evaluated three tasks across multiple LLMs with a controlled RTF baseline, it did not include adversarial inputs. Results demonstrate feasibility and structural superiority of the IR, but do not establish performance bounds for highly complex or adversarial data structures.

Fourth, the bridge map provides 365 legacy-to-modern entries, and the case study exercised 14 end-to-end on real data and 5 on the public benchmark. While this represents 19 of 365 entries at full end-to-end parity level, these 19 span the major clinical report categories (AE, demographics, disposition, ECG, laboratory, compliance, efficacy, survival). The remaining entries have been documented and validated at the unit level, making them already deployable in coexistence mode.

Fifth, the framework was designed for regulatory compliance (21 CFR Part~11 compatibility, audit trails, traceability) but has not undergone formal Installation Qualification, Operational Qualification, or Performance Qualification (IQ/OQ/PQ) validation. The coexistence mode adoption pattern (\secref{sec:arch:nda}) is intended to reduce, but not eliminate, this burden: the legacy library remains within its existing validation envelope and only the framework layer requires qualification.

Sixth, the software metrics used (LOC, parameter count, nesting depth, efferent coupling, cohesion) are established but indirect measures of software quality. Direct measures such as defect rates, mean time to resolution, and programmer productivity were not collected during this study.

\subsection{Future Work}

Immediate priorities include completing the full 365-entry parity matrix with end-to-end testing on both real-data and public-benchmark inputs, and formal IQ/OQ/PQ validation per Good Automated Manufacturing Practice (GAMP) 5 guidelines~[40]. Multi-site evaluation would test generalizability; we expect the coexistence-first adoption pattern to be the most directly transferable component. Integration with CDISC ARS could align the IR schema with an emerging industry standard. Cloud deployment of the SAS compute layer would enable scalable execution.

Expanding the polyglot interface beyond the current Python orchestrator and JSON IR endpoints is a natural next step. A native R client for the IR --- providing typed accessors over \code{jsonlite}-parsed cell grids and helpers for common cross-table consistency checks --- would make the framework directly consumable by the substantial population of clinical biostatisticians whose primary working language is R. Reciprocally, an R-based exploratory interface around the IR would complement the SAS-based regulated computation pathway and the Python-based AI/orchestration pathway, completing a three-language stack that addresses regulated computation, exploratory analytics, and AI integration through a single shared contract.

Expanding AI capabilities to cross-table QC, submission-level consistency validation, and interactive exploration of clinical results would further demonstrate the IR's practical value. A controlled comparison of LLM performance on IR versus RTF inputs across multiple models would provide stronger evidence for AI readiness claims.

\section{Conclusion}\label{sec:concl}

Large pharmaceutical organizations depend on legacy SAS macro libraries that have accumulated hundreds of components over decades of regulatory use. These libraries are difficult to integrate with cloud and AI workflows because they produce opaque formatted output with no structured intermediate representation, and they are difficult to modernize because every change to validated source code triggers re-validation cost. The conventional response --- a full rewrite onto a modern platform --- abandons decades of validated logic and is rarely affordable in practice.

This paper presented a systematic framework that resolves this dilemma by separating \emph{AI readiness} from \emph{source-level modernization}. The framework wraps the legacy library through a metadata layer comprising a 365-entry bridge map, a typed Intermediate Representation, and a Python orchestrator. As a result, the legacy macro library remains intact and within its existing validation envelope, yet its outputs are exposed as machine-readable JSON consumable by large language models, R analytics, and downstream automation. AI integration thereby becomes a Day-0 capability rather than the end state of a multi-year rewrite. Source-level consolidation, where an organization elects to undertake it, is offered as an opt-in continuous-improvement track governed by the same metadata, with the bridge map as the single deployment surface.

Applied to a 558-component industrial library, the framework was validated across 14 production report types from an internal Phase~III study (PROT008-SR1) and 5 report types from the public CDISC CDISCPilot01 benchmark. On the real-data track, 11 of 14 reports cleared the 80\% cell-level parity threshold (mean 82.7\%, best 99.2\%), with the remaining 3 having documented structural ceiling explanations rather than computational errors. On the public track, 5 reports achieved 100\% parity across 4{,}764 cells with 0 mismatches. A reusable taxonomy of twelve recurring divergence categories was identified; all 72 fixes were applied at the framework layer, demonstrating the fix-once / apply-many leverage of the IR contract. Where consolidation was also exercised, the modern core comprises 158 SAS macros (a 92\% reduction in lines of code) configured through 147 YAML files covering 39 report types --- but this reduction is reported as a measured upper bound on the optional consolidation pathway, not as a precondition for the AI-ready outcome.

The architectural separation produced a polyglot integration surface in addition to its consolidation potential. SAS retains exclusive responsibility for regulated computation; Python provides orchestration, IR-to-JSON serialization, the SAP parser, and the parity harness; R consumes the IR through standard JSON parsers for exploratory analytics; and large language models operate on the typed cell structure of the IR rather than on RTF. Each language addresses a single concern through one shared contract.

The pharmaceutical industry need not choose between preserving legacy systems and abandoning them. The framework described here offers a third path: keep the validated library, wrap it, and modernize at whatever rate the organization chooses. We encourage organizations maintaining large SAS macro libraries to evaluate this approach as a practical route toward AI-ready clinical reporting. Future work should focus on multi-site validation, native R client development, formal IQ/OQ/PQ qualification, and alignment with the CDISC Analysis Results Standard.

\section*{Author Contributions}
\addcontentsline{toc}{section}{Author Contributions}

JY conceptualised the framework, developed the architecture, conducted the case study, and wrote the manuscript.

\section*{Funding}
\addcontentsline{toc}{section}{Funding}

This research received no external funding.

\section*{Conflict of Interest}
\addcontentsline{toc}{section}{Conflict of Interest}

The author declares that the research was conducted in the absence of any commercial or financial relationships that could be construed as a potential conflict of interest.

\section*{Data Availability Statement}
\addcontentsline{toc}{section}{Data Availability Statement}

The CDISC CDISCPilot01 benchmark dataset used in this study is publicly available from the CDISC website (\url{https://github.com/cdisc-org/sdtm-adam-pilot-project}). The Intermediate Representation (IR) schema, the YAML configuration specification for 5 public-benchmark report types, and the parity validation harness used in the CDISCPilot01 evaluation are open-sourced and available on GitHub\footnote{Repository URL will be provided upon acceptance; currently anonymized for peer review.} to ensure full reproducibility. The AI readiness experiment prompts and structured IR JSON files used in Section~\ref{sec:ai} are included in the same repository. The modernized framework source code for the industrial case study (PROT008-SR1) is not publicly available due to organizational policy restrictions; however, the IR schema and evaluation methodology are designed to be framework-agnostic and applicable to any SAS macro library.


\end{document}